\documentclass[useAMS,usenatbib]{mn2e}

\usepackage{graphicx}
\usepackage{captcont}
\usepackage{times}

\def\aj{\,{AJ}}
\def\apj{\,{\rm ApJ}}

\def\apjs{\,{\rm ApJS}}

\def\pasj{\,{\rm PASJ}}

\def\aap{\,{\rm A\&A}}

\def\mnras{\,{\rm MNRAS}}

\title[E+A and Companion Galaxies - I]
{E+A and Companion Galaxies - I : A Catalogue and Statistics}
\author[Yamauchi, Yagi \& Goto]
{
  Chisato
  Yamauchi$^{1}$\thanks{E-mail:cyamauch@ir.isas.jaxa.jp}\thanks{Visiting Astronomer, Kitt Peak National Observatory, National Optical Astronomy Observatory, which is operated by the Association of Universities for Research in Astronomy, Inc. (AURA) under cooperative agreement with the National Science Foundation.},
  Masafumi Yagi$^{2}$\dag~ 
  and Tomotsugu Goto$^{2,3}$\dag 
  \\
  $^{1}$Institute of Space and Astronautical Science,
  Japan Aerospace Exploration Agency,
  3-1-1 Yoshinodai, Sagamihara, Kanagawa, 229-8510, Japan\\
  $^{2}$National Astronomical Observatory, 2-21-1 Osawa, Mitaka, Tokyo
  181-8588, Japan\\
$^{3}$Institute for Astronomy, University of Hawaii
2680 Woodlawn Drive, Honolulu, HI, 96822, USA\\
}
\begin{document}

\pagerange{\pageref{firstpage}--\pageref{lastpage}} \pubyear{2007}

\maketitle

\label{firstpage}

\begin{abstract}
Based on our intensive spectroscopic campaign with the GoldCam spectrograph
on the Kitt Peak National Observatory (KPNO) 2.1-m telescope,
we have constructed the first catalogue of E+A galaxies with spectroscopic companion galaxies,
and investigated a probability that an E+A galaxy have
close companion galaxies.
We selected 660 E+A galaxies with
$4.0{\rm \AA} < {\rm H}\delta~{\rm EW}$ at a redshift of $<0.167$
from the Data Release 5 of the Sloan Digital Sky Survey (SDSS).
 We selected their companion candidates from the SDSS imaging data,
and classified them into true companions,
 fore/background galaxies and companion candidates using the SDSS and our KPNO spectra.
We observed 
26
companion candidates of E+A galaxies at the KPNO
 to measure their redshifts.
Their spectra showed that 
17
targets are true companion galaxies.
The number of spectroscopically-confirmed E+A's companions are now 34.
This becomes the first catalogue of E+A galaxies with spectroscopic companion systems.
We found that 
E+A galaxies have an 54\% larger probability of having companion galaxies (7.88\%)
compared to the comparison sample of normal galaxies (5.12\%).
A statistical test shows the probabilities are different with 99.7\% significance.
Our results based on spectroscopy tightens the connection between the dynamical merger/interaction and the origin of E+A galaxies.
\end{abstract}

\begin{keywords}
galaxies: evolution, galaxies:interactions, galaxies:starburst, galaxies:peculiar
\end{keywords}

\section{Introduction}
\label{section:intro}

\citet{dre83,dre92} discovered galaxies with mysterious spectra 
in high-redshift clusters of galaxies.
These galaxies had strong Balmer absorption lines with no emission in
${\rm [OII]}$.  
They were named ``E+A'' galaxies
because their spectra resembled a superposition of those
of elliptical galaxies (${\rm Mg}_{5175}$, ${\rm Fe}_{5270}$ and ${\rm Ca}_{3934,3468}$
absorption lines) and 
A-type stars (Strong Balmer absorption)%
\footnote{Because the spectra of elliptical galaxies are characterised
by K stars, these galaxies are sometimes called
``k+a'' galaxies \citep[e.g.,][]{fra93,dre99,bar01}.
Following the first discovery, we refer to them as ``E+A'' throughout this paper.
}%
.
The existence of
strong Balmer absorption lines shows that these galaxies have
experienced starbursts within the last Gyr,
since the lifetime of an A-type star is about 1 Gyr.
However, they show no sign of ongoing star formation as 
indicated by non-detection in the ${\rm [OII]}$ emission line.  
Therefore, E+A galaxies are interpreted as post-starburst galaxies, 
i.e., galaxies that have undergone truncated starburst activity
\citep{dre83,dre92,cou87,mac88,new90,fab91,abr96}.
Thus, E+A galaxies have attracted a great deal of attention,
and some explanations for their origin have been proposed.

In the classical studies,
``E+A'' galaxies were found in cluster regions in both
low-redshift clusters \citep{fra93,cal93,cal97,cas01,ros01} and 
high-redshift clusters
\citep{sha85,lav86,cou87,bro88,fab91,bel95,bar96,fis98,mor98,cou98,dre99}.
Therefore, a cluster-specific phenomenon 
was thought to be responsible
for the violent star formation history of E+A galaxies.  
A ram-pressure stripping model
\citep{spi51,gun72,far80,ken81,aba99,fuj99,qui00,fuj04,fg04} 
as well as tides from the cluster potential \citep[e.g.,][]{fuj04}
may first
accelerate star formation of cluster of galaxies and later turn it off.
However, recent large surveys of the nearby Universe found many E+A
galaxies in the field regions \citep{got03,got03b,got05,bla04,qui04,hog06}.
It is obvious that these E+A galaxies in the field
regions cannot be explained by a physical mechanism that works in the
cluster region.  
E+A galaxies have often been thought to be transitional objects during
cluster galaxy evolution, involving phenomena such as the Butcher-Oemler
effect \citep[e.g.,][]{but78,rak95,mar01,ell01,kod01,got03a}, 
the morphology-density relation 
\citep[e.g.,][]{dre80,pos84,fas00,got03d,pos05,smi05}, 
and the correlation between various properties of the galaxies and the
environment \citep[e.g.,][]{tan04,pop07}.  
However, explaining cluster
galaxy evolution using E+A galaxies may no longer be realistic.

One explanation for E+A phenomena is dust-enshrouded star formation,
where E+A galaxies are actually star-forming, but emission
lines are invisible in optical wavelengths due to heavy obscuration
by dust \citep[e.g.,][]{pog00}.
A straightforward test for this scenario is observation in
radio wavelengths in which dust obscuration is negligible.
At 20-cm radio wavelengths, synchrotron radiation from electrons
accelerated by supernovae can be observed.
Therefore, in the absence of a radio-loud active nucleus, the radio flux
of a star-forming galaxy can be used to estimate its current massive
star formation rate (SFR) \citep{con92,ken98,hop03}. 
%
\citet{sma99} observed z=0.4 cluster in radio, 
and found that 5 out of 10 radio sources show 
E+A like spectra in optical wavelength.
\citet{cha01} observed 5 nearby field E+A galaxies
and detected no radio continuum.
\citet{mil01}
observed radio continua of 15 E+A galaxies and detected moderate levels
of star formation in only 2 of them.  
\citet{got04a} undertook 20-cm radio continuum observation of 36 E+A
galaxies and none of them were detected at 20-cm, 
suggesting that E+A galaxies are not dusty-starburst galaxies.


It is well-known that a
galaxy--galaxy interaction triggers off explosive star formation 
\citep[e.g.,][]{sch82,lav88,liu95a,liu95b,sch96,nik04}.  
\citet{oeg91} found a nearby E+A galaxy with a tidal tail feature.  
High-resolution Hubble Space Telescope imaging
supports the galaxy--galaxy interaction scenario
by showing some post-starburst (E+A) galaxies in high-redshift clusters
as having disturbed or interacting signatures
\citep{cou94,cou98,dre94,oem97}.  
\citet{nor01} performed long-slit spectroscopic observations of 21 E+A
galaxies, and found that young stellar populations of E+A galaxies are more
centrally concentrated than older populations, and old components of E+A
galaxies conform to the Faber--Jackson relation.  
\citet{bar01} reported that E+A galaxies, on average, tend to have
slightly bluer radial gradients toward the centre compared to normal
early-type galaxies.
\citet{yan04} presented HST observations of the five bluest E+A galaxies
with $z \sim 0.1$ and reported details of disturbed morphologies. 
Moreover, \citet{yan04}
detected compact sources associated with E+A galaxies consistent with
the brightest clusters in nearby starburst galaxies.  
\citet{yam05} not only showed obvious bluer radial colour gradients but
also found irregular structures in their 
two-dimensional (2-D) colour map.
%
%
%
\citet{yag06a} and 
\citet{yag06b} investigated the age distribution using
long-slit spectroscopy and found 
a positive gradient in the age of young stars from the centre to the
outer regions of the plume.
Recently, some numerical simulations on E+A galaxies have also
been presented.
\citet{bek01} modelled galaxy--galaxy mergers with dust extinction,
confirming that such systems can produce spectra that evolve into E+A
spectra.  
\citet{bek05} investigated the structural, kinematical and
spectrophotometric properties of E+A galaxies,
and showed that the 2-D distributions of line-of-sight
velocity, velocity dispersion, colour and line index in E+A galaxies
formed via the interaction and merging of two gas-rich spirals.

Thus, many studies have been conducted on the
large-scale environments and internal properties of E+A galaxies,
and recent studies basically support the merger/interaction origin of
E+A galaxies.
Several studies have focused on the 
medium-scale environments of E+A galaxies,
e.g., \citet{bla04}, 
\citet{got03}
and \citet{got05}.
Using SDSS imaging data,
\citet{got03} found that young E+A
galaxies have more accompanying galaxies within 100 kpc.
Results in \citet{got03} and \citet{got05} 
provide strong support for the merger/interaction origin
of E+A galaxies, and we are interested in the physical relation between 
E+A galaxies and their accompanying galaxies.
However, their studies are based on the imaging data, and
not all accompanying galaxies are spectroscopically observed in the
SDSS. 
In addition, it is unknown which galaxy is a real companion of E+A
galaxies in \citet{got03} and \citet{got05}.


Our aim is to extract some clues
from medium-scale environments, that is, the physical relations between 
E+A galaxies and their pair galaxies,
to investigate their evolution.
However, we must confirm
whether accompanying galaxies in \citet{got03} and \citet{got05}
are companions, using statistical analyses with spectroscopic
data.
We will begin our studies on the properties of
E+A galaxies and pair systems after such verification.

In this paper, we use publicly available 
{\it true} E+A galaxies (without ${\rm H}\alpha$ nor ${\rm [OII]}$ emission)
selected from the Sloan Digital Sky Survey 
(SDSS, \citealt[][]{yor00};
 Early Data Release, \citealt[][]{sto02}; 
 First Data Release, \citealt[][]{aba03};
 Second Data Release, \citealt[][]{aba04};
 Third Data Release, \citealt[][]{aba05};
 Fourth Data Release, \citealt[][]{ade06};
 Fifth Data Release, \citealt[][]{ade07}, hereafter DR5)
by \citet{got07b}.  
Both broadband imaging and the spectroscopic survey of 10,000 deg$^2$ of
SDSS provide us with the first opportunity to study a
very large number of E+A galaxies.
\citet{got07b} analysed $\sim$670,000 galaxy spectra
in the DR5, and the number of 
homogeneous E+A galaxies reached 1,062.
We created the catalogue of companions/candidates of E+A galaxies
using the SDSS SQL service, and we observed the spectra of
companion candidates to reveal their redshifts.
We then investigated an existing probability of companions of E+A
galaxies by a stricter analysis with spectroscopic data. 

This paper is organised as follows.
In section \ref{section:samples},
the definitions of two samples of 
galaxies are summarised.
In section \ref{section:observation},
we briefly describe our spectroscopic observations
on companion candidates of E+A galaxies, and
reduction of their data.
In section \ref{section:method}, we explain a
somewhat complicated method of statistical analysis 
for readers to more easily understand.
In section \ref{section:analysis_and_results}, we show
the application of the method described in section 
\ref{section:method} to both the E+A sample and control sample,
and present the results.
Lastly, we provide a discussion and summary in
section \ref{section:discussion}.

Unless otherwise stated, we adopt the best-fitting 
{\it Wilkinson Microwave Anisotropy Probe (WMAP)}
cosmology: $(h,\Omega_m,\Omega_L) = (0.71,0.27,0.73)$
\citep{ben03,kom08}.


\begin{figure*}
 \begin{center}
  \includegraphics[scale=0.30]{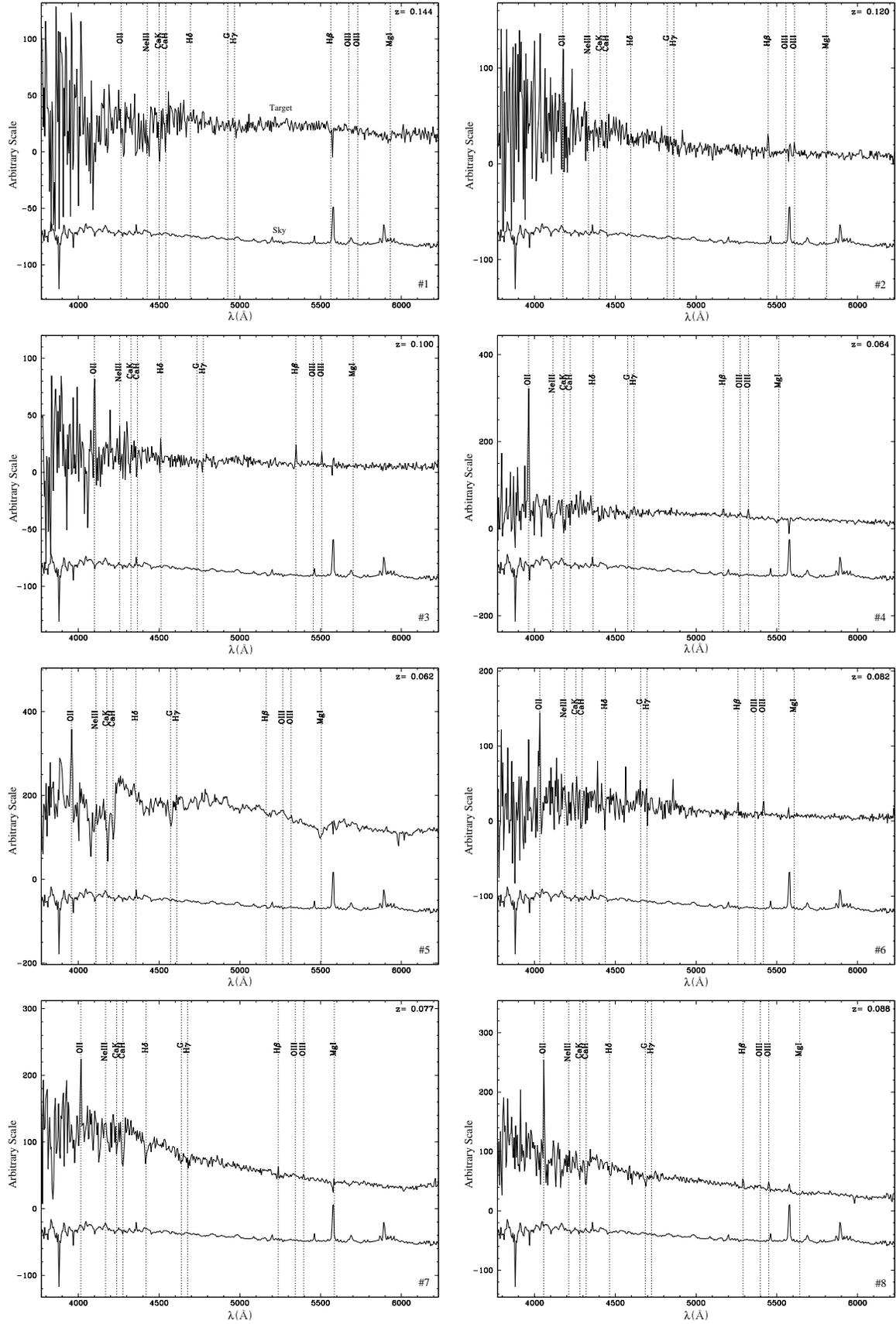}
 \end{center}

 \caption{Spectra of 16 new E+A's companions taken
 with the KPNO 2.1-m telescope applying a 20\AA~binning. 
 A sky spectrum in the observed data is shown at the bottom of each
 panel.
 This figure includes
 targets \#1({\it top left}), \#2({\it top right}), ..., 
 \#7({\it bottom left}) and \#8({\it bottom right}).
 }\label{fig:spectra}
\end{figure*}

\begin{figure*}
 \begin{center}
  \includegraphics[scale=0.30]{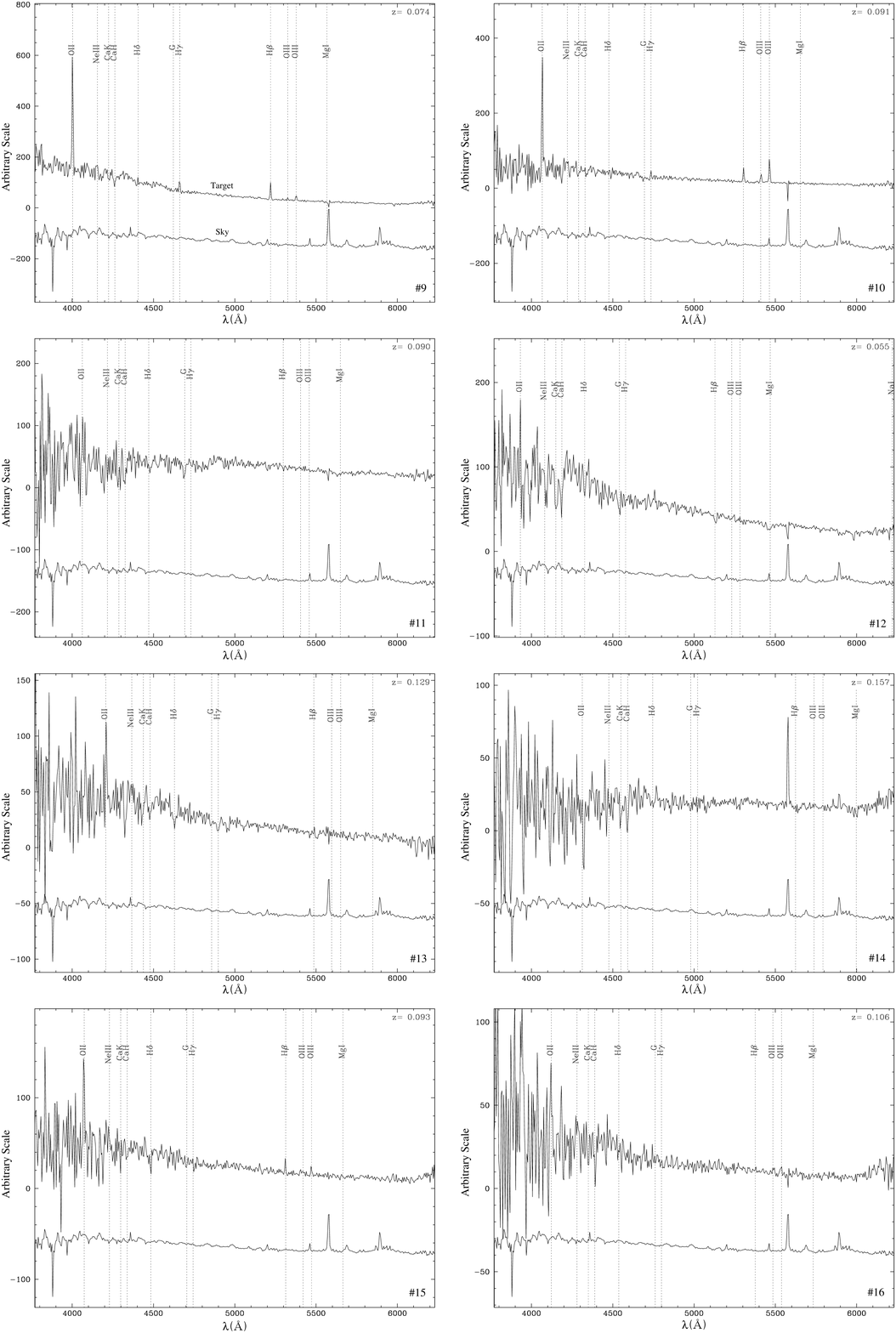}
 \end{center}
 \contcaption{
 targets \#9({\it top left}), \#10({\it top right}), ..., 
 \#15({\it bottom left}) and \#16({\it bottom right}).
 }
\end{figure*}

\begin{figure*}
 \begin{center}
  \includegraphics[scale=0.30,angle=-90]{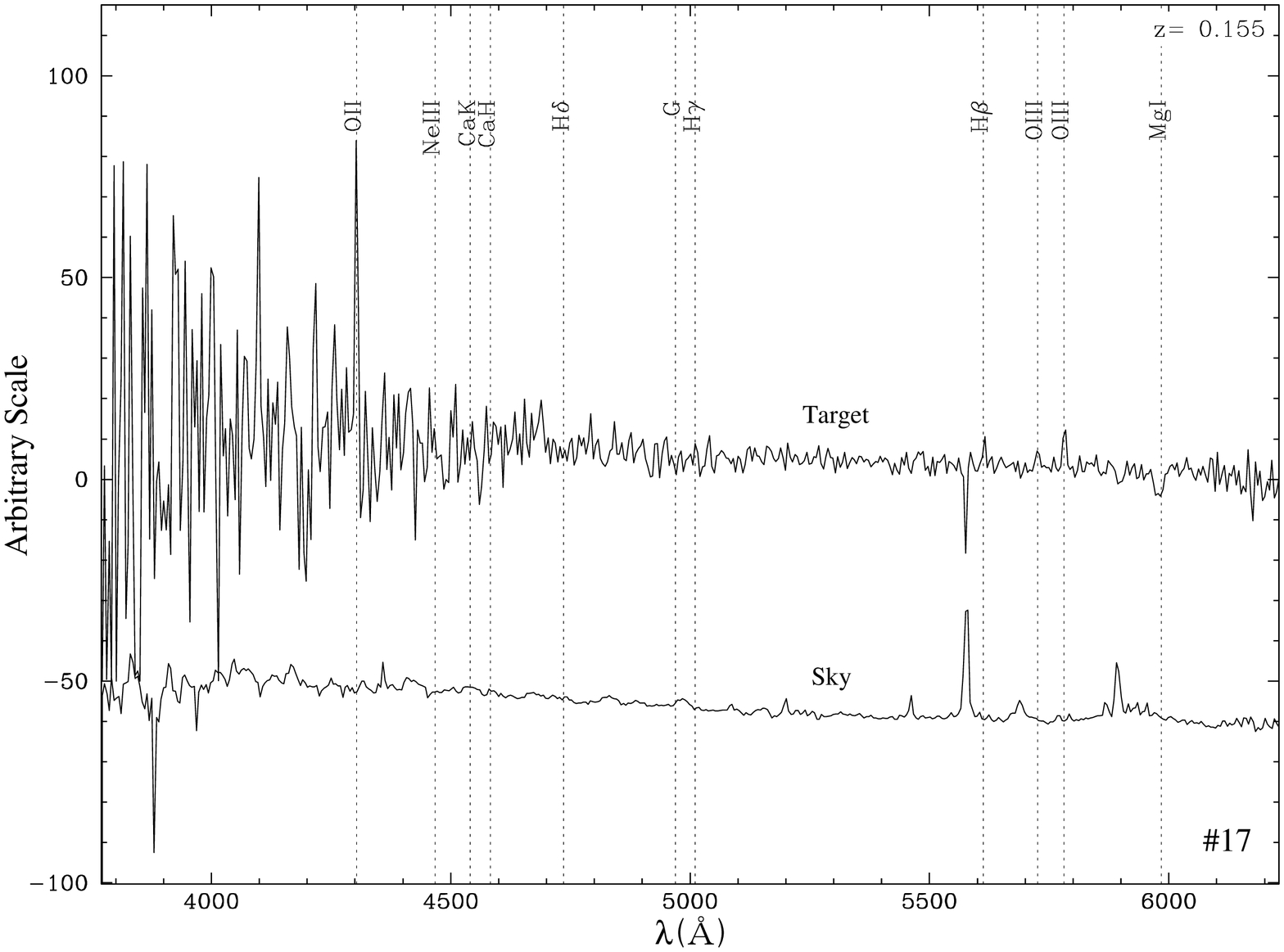}
 \end{center}
 \contcaption{
 target \#17.
 }
\end{figure*}

\begin{table*}
\begin{center}
\caption{List of E+A and companion pairs.
 Columns $r$ and $M_r$
 are magnitudes and absolute magnitudes with
 reddening- and {\it K}-correction by {\tt kcorrect.v4\_1\_4} \citep{bla03}.
 Column elements of the ``origin of $z$'' filled by
 italicised letters show information obtained by the NED search service.
 ID 25 or ID 26 is an ``E+A and E+A'' pair.
 }\label{table:all_list} {\tabcolsep=0.9mm
\begin{tabular}{cccccccccccccccc}
\hline
\hline
\multicolumn{1}{c}{Pair} &
\multicolumn{6}{c}{E+A galaxies} &
\multicolumn{5}{c}{Companion galaxies} &
\multicolumn{1}{c}{} &
\multicolumn{1}{c}{~} &
\multicolumn{1}{c}{\scriptsize projection} \\
\multicolumn{1}{c}{ID} &
\multicolumn{1}{c}{R.A.} &
\multicolumn{1}{c}{Dec.} &
\multicolumn{1}{c}{$r$} &
\multicolumn{1}{c}{$M_r$} &
\multicolumn{1}{c}{$z$} &
\multicolumn{1}{c}{\scriptsize ${\rm H}\delta~{\rm EW}$} &
\multicolumn{1}{c}{R.A.} &
\multicolumn{1}{c}{Dec.} &
\multicolumn{1}{c}{$r$} &
\multicolumn{1}{c}{$M_r$} &
\multicolumn{1}{c}{$z$} &
\multicolumn{1}{c}{origin of $z$} &
\multicolumn{1}{c}{$|\Delta z|$} &
\multicolumn{1}{c}{\scriptsize distance(kpc)} \\
\hline
 1 & 
 01:19:42.23 & +01:07:51.5 & 16.75 & -21.29 & 0.090 & 6.40 &
 01:19:41.58 & +01:07:39.9 & 17.63 & -20.41 & 0.089 &
 SDSS & 0.001 & 25.18 \\
 2 &
 07:53:23.09 & +24:33:00.5 & 15.80 & -21.36 & 0.061 & 5.74 &
 07:53:22.80 & +24:33:02.5 & 15.78 & -21.38 & 0.062 &
 SDSS & 0.001 & 5.21 \\
 3 &
 08:41:41.13 & +26:42:39.2 & 16.97 & -21.07 & 0.090 & 4.97 &
 08:41:40.75 & +26:42:41.2 & 17.23 & -20.81 & 0.090 &
 KPNO Observation \#11 & 0.000 & 9.11 \\
 4 &
 09:03:32.77 & +01:12:36.4 & 16.16 & -20.89 & 0.058 & 4.43 &
 09:03:32.99 & +01:12:31.7 & 17.39 & -19.66 & 0.058 &
 SDSS & 0.000 & 6.32 \\
 5 &
 09:25:03.81 & +40:46:02.8 & 17.69 & -20.97 & 0.117 & 4.90 &
 09:25:04.65 & +40:45:55.3 & 16.53 & -22.12 & 0.117 & 
 SDSS & 0.000 & 25.25 \\
 6 &
 09:58:24.62 & +00:32:03.1  & 17.15 & -20.96 & 0.093 & 5.50 &
 09:58:24.08 & +00:32:06.4  & 17.45 & -20.66 & 0.093 &
 KPNO Observation \#15 & 0.000 & 14.89 \\
 7 & 
 10:05:22.45 & +07:27:28.3  & 16.95 & -20.28 & 0.063 & 4.92 &
 10:05:21.12 & +07:27:25.9  & 15.38 & -21.85 & 0.064 &
 KPNO Observation \#4 & 0.001 & 23.99 \\
 8 &
 10:16:29.22 & -00:01:37.1 & 17.37 & -21.02 & 0.105 & 5.90 &
 10:16:28.55 & -00:01:34.3 & 17.86 & -20.54 & 0.105 & 
 {\it MGC 0004420} & 0.000 & 19.43 \\
 9 &
 10:57:49.08 & +41:00:35.9 & 17.50 & -21.38 & 0.128 & 4.05 &
 10:57:47.54 & +41:00:49.2 & 17.16 & -21.72 & 0.128 & 
 SDSS & 0.000 & 49.89 \\
 10 &
 11:10:04.24 & +11:51:19.7  & 15.45 & -22.25 & 0.078 & 4.13 &
 11:10:03.74 & +11:51:23.1  & 17.10 & -20.61 & 0.077 &
 KPNO Observation \#7 & 0.001 & 11.78 \\
 11 &
 11:41:38.76 & +09:43:44.6 & 17.64 & -20.07 & 0.078 & 4.64 &
 11:41:36.84 & +09:43:35.6 & 15.46 & -22.25 & 0.079 & 
 SDSS & 0.001 & 43.52 \\
 12 &
 11:52:08.57 & +48:16:58.4  & 17.24 & -19.67 & 0.055 & 5.88 &
 11:52:07.36 & +48:17:22.7  & 16.66 & -20.25 & 0.055 &
 KPNO Observation \#12 & 0.000 & 28.52 \\
 13 &
 11:56:28.91 & +48:55:41.6 & 17.72 & -19.99 & 0.078 & 4.06 &
 11:56:27.66 & +48:55:39.9 & 16.51 & -21.20 & 0.078 & 
 SDSS & 0.000 & 18.14 \\
 14 &
 12:13:33.02 & +14:29:00.1 & 16.26 & -21.01 & 0.064 & 5.94 &
 12:13:34.12 & +14:28:42.4 & 15.94 & -21.33 & 0.064 &
 SDSS & 0.000 & 28.90 \\
 15 &
 12:22:40.46 & +15:02:05.4 & 17.72 & -19.82 & 0.072 & 8.10 &
 12:22:40.76 & +15:02:22.3 & 15.67 & -21.88 & 0.072 &
 SDSS & 0.000 & 23.73 \\
 16 &
 13:00:29.45 & +54:55:04.0  & 16.60 & -21.41 & 0.089 & 4.81 &
 13:00:32.15 & +54:54:57.9  & 16.65 & -21.36 & 0.088 &
 KPNO Observation \#8 & 0.001 & 39.44 \\
 17 &
 13:30:24.73 & +02:23:25.8 & 17.33 & -20.41 & 0.079 & 5.66 &
 13:30:23.85 & +02:23:04.3 & 15.67 & -22.06 & 0.079 & 
 SDSS & 0.000 & 37.22 \\
 18 &
 13:50:05.58 & -02:47:34.7  & 16.71 & -22.16 & 0.128 & 5.81 &
 13:50:05.89 & -02:47:38.8  & 17.94 & -20.93 & 0.129 &
 KPNO Observation \#13 & 0.001 & 14.02 \\
 19 &
 13:50:50.98 & +02:19:38.4 & 15.85 & -19.90 & 0.033 & 4.59 &
 13:50:53.55 & +02:19:24.6 & 14.28 & -21.46 & 0.033 & 
 {\it UGC 08750 NED01} & 0.000 & 26.25 \\
 20 &
 14:24:53.14 & +23:07:46.8  & 15.70 & -21.89 & 0.074 & 4.05 &
 14:24:52.86 & +23:07:20.9  & 16.82 & -20.77 & 0.074 &
 KPNO Observation \#9 & 0.000 & 36.24 \\
 21 &
 14:36:33.54 & +55:53:17.8 & 17.60 & -20.78 & 0.104 & 4.08 &
 14:36:31.82 & +55:53:27.6 & 16.65 & -21.74 & 0.105 & 
 SDSS & 0.001 & 32.96 \\
 22 &
 14:50:33.38 & +03:18:20.0 & 17.33 & -19.86 & 0.062 & 4.04 &
 14:50:31.09 & +03:18:03.5 & 16.89 & -20.30 & 0.062 &
 SDSS & 0.000 & 44.90 \\
 23 &
 15:13:09.04 & +33:49:44.4  & 17.47 & -21.84 & 0.155 & 5.05 &
 15:13:07.74 & +33:49:51.4  & 19.48 & -19.83 & 0.155 &
 KPNO Observation \#17 & 0.000 & 46.63 \\
 24 &
 15:34:16.00 & +03:59:34.6  & 17.71 & -21.65 & 0.157 & 4.55 &
 15:34:16.19 & +03:59:33.4  & 18.32 & -21.03 & 0.157 &
 KPNO Observation \#14 & 0.000 & 8.27 \\
 25 &
 15:57:28.87 & +27:25:45.2 & 16.43 & -21.55 & 0.087 & 4.22 &
 15:57:30.42 & +27:25:41.1 & 17.30 & -20.68 & 0.088 & 
 SDSS & 0.001 & 34.10 \\
 26 &
 15:57:30.42 & +27:25:41.1 & 17.30 & -20.69 & 0.088 & 4.11 &
 15:57:28.87 & +27:25:45.2 & 16.43 & -21.56 & 0.087 &
 SDSS & 0.001 & 34.23 \\
 27 & 
 16:13:30.18 & +51:03:35.5 & 15.48 & -20.34 & 0.034 & 7.57 &
 16:13:32.23 & +51:03:42.9 & 15.17 & -20.65 & 0.033 & 
 {\it I Zw 136 NOTES02} & 0.001 & 13.64 \\
 28 &
 16:21:51.96 & +49:28:59.8  & 17.04 & -20.79 & 0.082 & 4.26 &
 16:21:55.07 & +49:29:08.0  & 18.21 & -19.62 & 0.082 &
 KPNO Observation \#6 & 0.000 & 47.84 \\
 29 &
 16:23:01.30 & +23:00:39.8  & 17.48 & -19.69 & 0.061 & 4.73 &
 16:23:01.04 & +23:01:12.0  & 14.89 & -22.28 & 0.062 & 
 KPNO Observation \#5 & 0.001 & 37.84 \\
 30 &
 16:23:17.62 & +32:45:26.8  & 17.73 & -20.32 & 0.090 & 4.65 &
 16:23:17.42 & +32:45:02.3  & 17.24 & -20.81 & 0.091 & 
 KPNO Observation \#10 & 0.001 & 40.97 \\
 31 &
 16:39:25.01 & +30:37:09.8  & 17.56 & -20.86 & 0.106 & 8.23 &
 16:39:24.87 & +30:37:15.3  & 18.00 & -20.42 & 0.106 &
 KPNO Observation \#16 & 0.000 & 11.09 \\
 32 & 
 16:56:48.64 & +31:47:02.3 & 16.70 & -21.58 & 0.100 & 5.75 &
 16:56:48.32 & +31:47:02.3 & 18.63 & -19.66 & 0.100 &
 KPNO Observation \#3 & 0.000 & 13.39 \\
 33 & 
 17:03:56.71 & +62:28:48.3 & 17.64 & -21.53 & 0.145 & 7.29 &
 17:03:55.12 & +62:28:56.0 & 18.03 & -21.14 & 0.144 & 
 KPNO Observation \#1 & 0.001 & 33.89 \\
 34 & 
 17:08:59.24 & +32:20:53.1 & 17.58 & -21.14 & 0.121 & 6.03 &
 17:08:59.03 & +32:21:00.3 & 18.08 & -20.65 & 0.120 &
 KPNO Observation \#2 & 0.001 & 16.49 \\

\hline
\end{tabular} 
}
\end{center}
\end{table*}


\section{Samples}
\label{section:samples}

Our statistical analysis requires E+A and control samples
and their companions/candidates,
and the redshifts of some candidates are measured by our observations.
We describe our method for creating these samples in this section.

Galaxies used in our study are taken from the SDSS DR5 \citep{ade07}.
Details of the photometric system, imaging hardware and astrometric calibration of
the SDSS are described 
in \citet{fuk96,gun98,hog01,smi02,str02,pie03}.
Our E+A galaxies are selected from a publicly available catalogue
described in \cite{got07b}.
We created a catalogue of companions/candidates of E+A galaxies
and a comparison sample of normal galaxies.
These samples satisfy an absolute magnitude of 
$-22.5 < M_r < -19.5$ ($k$- and Galactic extinction corrected) using the SDSS Catalogue Archive Server (CAS).
This absolute magnitude range is 
adopted
so that the control sample of normal galaxies should be volume-limited. 

\subsection{Parent Galaxies}

The SDSS DR5 catalogue contains $\sim$670,000 galaxies with spectroscopic information.
\citet{got07b} selected 1062 E+A galaxies which satisfy
$4.0{\rm \AA}<{\rm H}\delta~{\rm EW}$,
$-3.0{\rm \AA}<{\rm H}\alpha~{\rm EW}$ and
$-2.5{\rm \AA}<{\rm [OII]}~{\rm EW}$ (absorption lines have a positive
sign).
The E+A galaxies are selected in an unbiased way 
except for the redshift cut ($z>0.032$) and the S/N cut 
($>$10 per pixel in $r$-band wavelength).
The redshift cut guarantees a reliable measurement of 
${\rm [OII]}$, since the blue limit of the SDSS spectrum is 3800\AA.
The spectroscopic data, ${\rm H}\delta$, ${\rm H}\alpha$ and ${\rm [OII]}$
equivalent width (${\rm EWs}$) and their errors are measured by the 
flux-summing method described in \citet{got07b}
(See also \citealt[][]{got03}).

Our targets have an absolute magnitude of $-22.5 < M_r < -19.5$ and $z<0.167$.
These criteria assure all targets are 
brighter than $r=20.0$, and thus, can be observed spectroscopically with
the KPNO 2.1-m telescope. 
This selection leaves 660 E+A {\em parent}%
\footnote{Note that the magnitude or physical size of the `parent galaxy'
is not always larger than that of `companion galaxies'
in this paper.
}
galaxies around which we spectroscopically look for companion galaxies. We use these
as our 
sample for statistical study.

A control sample is selected from all objects classified as galaxies in
the SDSS DR5 catalogue.
The magnitude range of the SDSS spectroscopy 
that satisfies high completeness and reliability is
$14.5 < r < 17.7$.  We obtain
$r$-band distance modulus $(m-M)_r \sim 37$
at $z=0.06$.
Therefore, we can set the limit of redshift ($z$)
between 0.0570 and 0.0620
so that the sample satisfies the volume-limited condition
with $-22.5 < M_r < -19.5$; 
then we obtain 11,267 {\em parent} normal galaxies as a control sample.

The control sample and the E+A sample are selected from the
different redshift ranges ($0.032<z<0.167$ and $0.0570<z<0.0620$, respectively).
However, both of the redshift ranges are small and close to $z=0$, and
thus, we assume that evolutionary differences 
such as change of merger rate between two samples are small enough in
our samples.


\subsection{Companion/Candidate Galaxies}
\label{section:companion/candidate}

To investigate the number of parent galaxies that have companion
galaxies,
we created a catalogue of companion candidates 
for our E+A sample and control sample
using a photometric catalogue.
The companion candidates are selected within the region of 50-kpc radius
centred on the parent galaxy in the sky.
The candidate catalogue is constructed by
executing the {\tt fGetNearbyObjEq()} function built in the SDSS SQL
service for each parent galaxy.
We calculated the absolute magnitude $M_r$ of candidates assuming
that they are placed at the redshift of the parent galaxy, and then
removed candidates whose $M_r$ 
do not satisfy 
$-22.5 < M_r < -19.5$.
The numbers of parent E+A and normal galaxies that have companions or
candidates are 97/660 and 791/11267, respectively.

The next step is to distinguish the true companion galaxies from the candidates.
We obtained the redshifts of the candidate galaxies that have 
{\tt SpecObjID} from the SDSS SQL server, and selected the true companions.
 In addition to this, we utilized the redshift information in the NED database whenever available.
We used the criterion of $|z| \leq 0.002$ to decide whether
a candidate galaxy is a true companion.
The escape velocity of a $M_r=-22.5$ galaxy is $\sim 200$km/s,
when $M_{\odot}/L_r \simeq 3$ \citep{kau03} is assumed.  
Therefore, the criteria of $|z| = 0.002$,
that is, $v \simeq 600$km/s, is large enough not to miss any 
gravitationally bound galaxies.
We use the notation `companions'
for the galaxies that satisfy the criteria
throughout this paper.

\section{Observation with the KPNO 2.1-m telescope}
\label{section:observation}

There are many 
companion candidates
without spectroscopic redshift
of E+A galaxies in our catalogue created 
in section \ref{section:companion/candidate}.
We used the KPNO 2.1-m telescope to obtain 
some of their spectroscopic redshifts.

We used the GoldCam spectrometer attached to
the
2.1-m telescope at the KPNO.
The aim of this observation was to identify 
the true companions of E+A galaxies. 
This purpose can be achieved by obtaining spectra of
accompanying galaxies from the SDSS photometric catalogue
since we already know the redshifts of parent E+A galaxies.
We used the 26new grating 
with the long-slit of 2 $\times$ 300 arcsec.
The CCD for the spectroscopy is 3,072$\times$1,024 pixel
whose resolution is 0.78 arcsecs/pixel and $\sim$ 5\AA.
We used a quartz lamp and HeNeAr calibration source built in the GoldCam
for flat and comparison, respectively.


The targets were selected from 
companion candidates
without spectroscopic redshifts.
Our observation was carried out in three semesters.
We observed 
26
companion candidates
\footnote{Actually we observed 28 candidates in total,
 including two backup targets. See APPENDIX1.}
on
September 22-24, 2005, June 28, 2006 and
May 17-21, 2007.
The data reduction were performed with the NOAO/IRAF
V2.12.2.
The {\tt ccdproc} task in {\tt noao/imred/ccred} package was used
for the overscan reduction and zero, dark and flat corrections.
The comparison, zero, dark and target frames
were simply combined by the {\tt imcombine}, but
the flat frames were combined after being
divided by the mean value calculated by the {\tt imstat} task.

The wavelength calibration was performed 
by {\tt identify}, {\tt reidentify} and {\tt fitcoords}
tasks using a HeNeAr comparison data.
The {\tt transform} task transforms long-slit images to wavelength
co-ordinates using the calibration database.
After carefully subtracting the background by the {\tt background} task,
we traced the signal of the target frame with a 4-pixel aperture
by the {\tt apall} task.

Our companion candidates included some passive galaxies and galaxies
fainter than $r=19$ magnitude.
The S/N of such objects cannot but be poor (S/N$\sim$1) when
using the 2.1-m telescope: therefore, sometimes it was
difficult 
to identify their redshifts.

In summary, we identified the redshifts of 
19 targets
and could not identify those of 
7
targets.
We found that these 
galaxies
with spectroscopic redshift
included
17 
true companions.
We show their spectra 
in Figure \ref{fig:spectra}.
Table \ref{table:all_list} also includes basic data of
these successful targets.


\section{Basic Method of Statistical Analysis}
\label{section:method}

\setlength{\tabcolsep}{1pt}


The goal of our study is to compare the fractions
of galaxies that have companions 
within a defined region (e.g., within a 50-kpc radius) in the sky,
for E+A and control samples.
If the redshift of all galaxies are already known,
we can easily count the number of parent galaxies with companions.
However, many galaxies have no spectroscopic data, 
and we cannot know whether the galaxy is a companion or a chance overlap.
Therefore, we have to estimate the fraction of galaxies having companions
using statistical analysis.
We describe the analytical method in this section.

\begin{figure}
\includegraphics[scale=0.65]{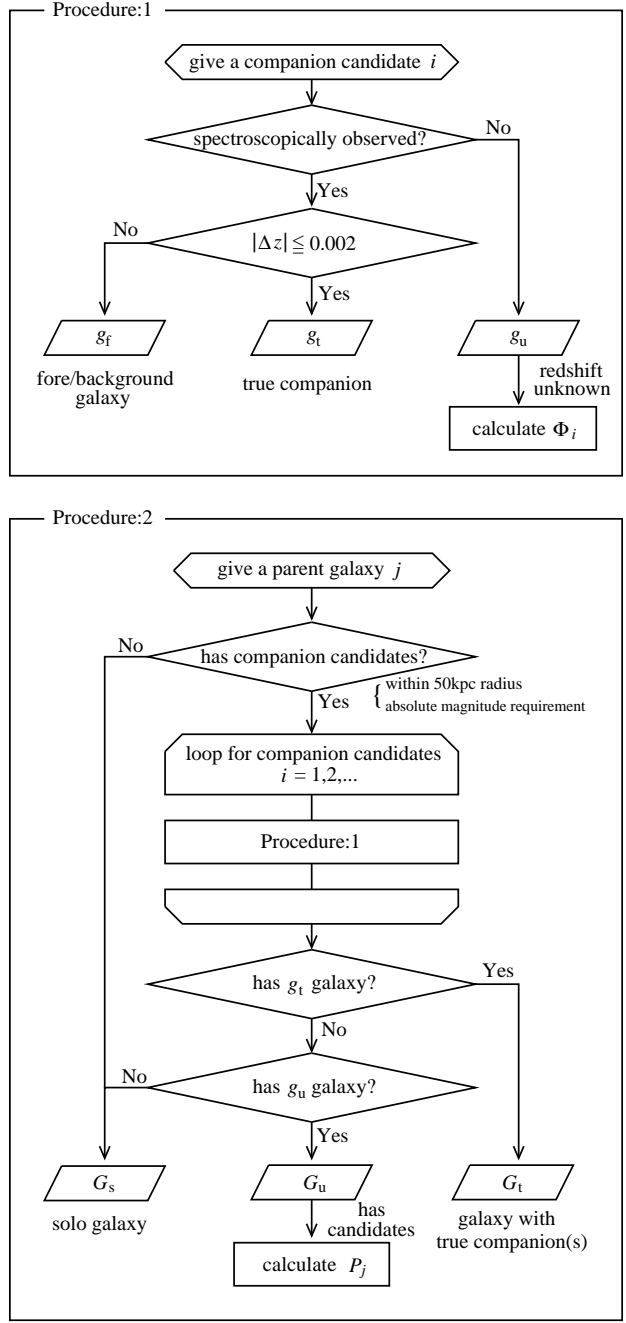}
 \caption{Basic flow of
 our analysis.  Top panel `procedure:1' shows
 the flow about a given companion candidate,
 and the bottom panel `procedure:2' shows
 that about a given parent galaxy.
 `Procedure:1' is used in `procedure:2'.
 }\label{fig:flow}
\end{figure}


\subsection{Estimation of the number of companion galaxies within a defined region}

The companion candidates in the constructed catalogue can be
divided into following groups:\\
\begin{tabular}{lll}
 $g_{\rm t}$ & : & companions with spectroscopic redshifts,\\
 $g_{\rm f}$ & : & fore/background galaxies with spectroscopic redshifts,\\
 $g_{\rm u}$ & : & companion candidates without spectroscopic redshifts.\\
\end{tabular}
\\
We define $n_{\rm t}$, $n_{\rm f}$ and $n_{\rm u}$ as the number of 
$g_{\rm t}$, $g_{\rm f}$ and $g_{\rm u}$ galaxies, respectively.
The following notations are also defined:\\
\begin{tabular}{lll}
$n_{\rm os}$ & : & number of accompanying galaxies with spectroscopic data \\
$n_{\rm op}$ & : & number of accompanying galaxies with photometric data\\
\end{tabular}\\
That is, $n_{\rm os} = n_{\rm t} + n_{\rm f}$ and
$n_{\rm op} = n_{\rm t} + n_{\rm f} + n_{\rm u}$.
The actual flow is presented in the top panel in Figure \ref{fig:flow};
the problem is the number of true companions in $g_{\rm u}$.

The number of {\em all} companion galaxies, 
$n_{\rm et}$, 
is the sum of the number of true companions ($n_{\rm t}$)
and the number statistically estimated,
which is a part of $n_{\rm u}$.
We assume that the target selection of SDSS spectroscopy
is unbiased \citep{str02}. 
Under the assumption, the existing probability 
of companions in the photometric sample equals the probability in
the spectroscopic sample.
We can derive 
\begin{equation}
\label{eq:n_et}
 \frac{n_{\rm et}}{n_{\rm op}} = \frac{n_{\rm t}}{n_{\rm os}}.
\end{equation}
The expected number of companions 
per a parent galaxy
within a defined region,
$\Phi_i$, is therefore obtained by
\begin{equation}
\label{eq:Phi}
 \Phi_i= \frac{n_{\rm et}}{N_{\rm M}},
\end{equation}
where $N_{\rm M}$ is the number of parent galaxies.
In this study, we use lowercase $g$ and $n$ for
group and number of companion/candidate galaxies, and 
uppercase $G$ and $N$ for those of parent galaxies.

\subsection{Estimation of number of parent galaxies that have companions}

The parent galaxies are also divided into three groups:\\
\begin{tabular}{lll}
$G_{\rm t}$&:& galaxies that have true companions 
(may also have candidates\\
&& without redshift or fore/background galaxies),\\
$G_{\rm s}$&:& solo galaxies (around which there may be fore/background\\
&& galaxies with spectroscopic redshifts), \\
$G_{\rm u}$&:& galaxies that have no true companions but have candidates\\
&& without spectroscopy (around which there may also be \\
&& fore/background galaxies with spectroscopic redshifts) \\
\end{tabular}\\
The targets for spectroscopic observation at the KPNO are selected 
from the $G_{\rm u}$ group.
Note that we selected the parent galaxies from the spectroscopic catalogue,
and we therefore know the redshifts of the parent galaxies.
We show the overview of this grouping procedure
in the bottom panel in Figure \ref{fig:flow}.
Our goal is to obtain the number of $G_{\rm t}$ plus 
the number of a portion of $G_{\rm u}$, 
since a companion candidate without spectroscopy
is either `a true companion' or `a back/foreground galaxy'.
The problem is the estimation of the probability of
existing companions in $G_{\rm u}$ galaxies.

A companion candidate without spectroscopy
is either `a true companion' or `a back/foreground galaxy'.
The probability that a candidate $i$ is 
a true companion, $p_i$, can be calculated by
%
%
%
\begin{equation}
\label{eq:p}
 p_i = \Phi_i/\phi_i ,
\end{equation}
where $\phi_i$ is the expected number of galaxies within
the corresponding field of view (FOV) of a certain physical area 
in which the parent galaxy is centrally placed.

%
%

Allocating the probability $p_i$ to each companion candidate $i$
of the parent galaxies in $G_{\rm u}$, 
we can calculate the probability
that the parent galaxy $j$ has at least one true companion by
\begin{equation}
\label{eq:P}
P_j= 1-\prod_{i} (1-p_i) .
\end{equation}
In group $G_{\rm u}$, the expected number of galaxies 
that have companion galaxies, $N_{\rm e}$, is the summation 
of $P_j$ of each parent galaxy $j$,
\begin{equation}
\label{eq:N_e}
N_{\rm e} = \sum_{j}^{N_{\rm u}} P_j .
\end{equation}
We can thus estimate the number of parent galaxies with
companion galaxies as
\begin{equation}
\label{eq:N_et}
 N_{\rm et} = N_{\rm t} + N_{\rm e} ,
\end{equation}
where $N_{\rm t}$ is the number of galaxies in $G_{\rm t}$.


\begin{table}
\begin{center}
\caption{Number counts of galaxies 
 $\nu_{50}$ (per 0.5 mag deg$^{-2}$) in
 the $r$-band
within the corresponding FOVs of 50-kpc radius
for 5 redshift regions.
$N_{\rm p}$ is the number of parent galaxies
centrally placed in the FOVs, and
$\sum n_{\rm f}$ is the number of found galaxies
in the FOVs in total.
 }\label{table:ncount_list} {\tabcolsep=0.55mm
\begin{tabular}{cccccc}
\hline
\hline
\multicolumn{1}{c}{\scriptsize Mag Range ($r$)} &
\multicolumn{1}{c}{\footnotesize $z$:0.03-0.06} &
\multicolumn{1}{c}{\footnotesize $z$:0.06-0.09} &
\multicolumn{1}{c}{\footnotesize $z$:0.09-0.12} &
\multicolumn{1}{c}{\footnotesize $z$:0.12-0.15} &
\multicolumn{1}{c}{\footnotesize $z$:0.15-0.18} 
~\\
\hline
 13.0-13.5  & 0.550  & 0.408 &  0.0    & 0.0   & 0.0   \\
 13.5-14.0  & 2.31   & 0.316 &  0.688  & 2.42  & 0.0   \\
 14.0-14.5  & 5.00   & 0.836 &  0.539  & 0.0   & 0.0   \\
 14.5-15.0  & 9.91   & 4.21  &  0.947  & 0.939 & 0.0   \\
 15.0-15.5  & 14.7   & 13.9  &  6.279  & 6.230 & 2.03  \\
 15.5-16.0  & 17.9   & 23.6  &  21.2   & 5.345 & 6.53  \\
 16.0-16.5  & 30.3   & 34.8  &  42.7   & 28.6  & 17.4  \\
 16.5-17.0  & 36.1   & 47.3  &  75.8   & 68.9  & 54.0  \\
 17.0-17.5  & 49.6   & 70.8  &  87.0   & 111.4 & 114.2 \\
 17.5-18.0  & 83.2   & 103.8 &  158.2  & 189.7 & 265.8 \\
 18.0-18.5  & 151.2  & 176.3 &  230.6  & 366.8 & 430.2 \\
 18.5-19.0  & 227.5  & 263.0 &  359.1  & 480.3 & 645.8 \\
 19.0-19.5  & 378.1  & 389.3 &  528.5  & 626.1 & 864.3 \\
 19.5-20.0  & 605.8  & 676.1 &  752.3  & 975.1 & 1211  \\
 20.0-20.5  & 944.7  & 1010  &  1242   & 1419  & 1714  \\
 20.5-21.0  & 1415   & 1493  &  1698   & 1966  & 2399  \\
 21.0-21.5  & 2032   & 2181  &  2395   & 2734  & 3146  \\
\hline
 $N_{\rm p}$ & 5807 & 12481 & 10061 & 8538 & 5782 \\
 $\sum n_{\rm f}$  & 27267  & 25447  &  12915   & 8382  & 4949 \\
\hline
\end{tabular} 
}
\end{center}
\end{table}

\begin{figure}
\includegraphics[scale=0.59]{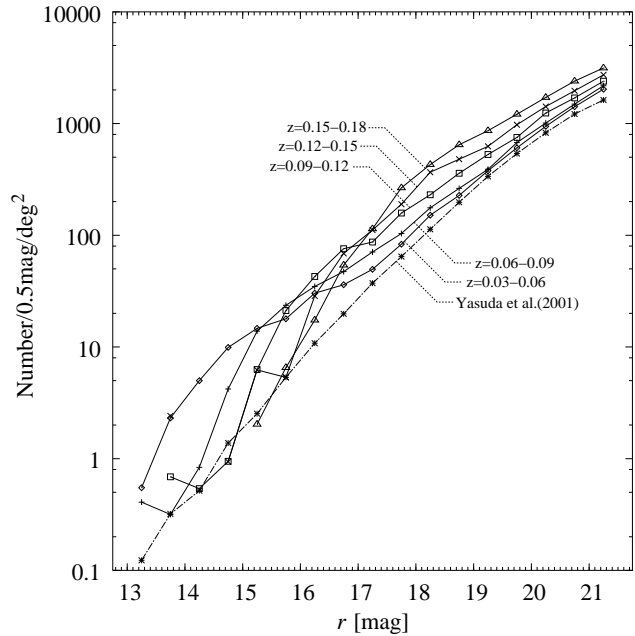}
\caption{
Plot of number count in Table \ref{table:ncount_list}
and the result of \citet{yas01}.
Our number count of 5 redshift regions
are drawn by solid line, and
the dot-dashed line is the count of \citet{yas01}.
}\label{fig:ncount}
\end{figure}

\section{Analysis and Results}
\label{section:analysis_and_results}


Using the method described in section \ref{section:method},
we estimated the number of companion galaxies of 
control sample and E+A sample.
As described in section \ref{section:samples},
the control sample is made from SDSS DR5 with
$0.0570 < z < 0.0625$ and $-22.5 < M_r < -19.5$.
The E+A sample is made from \citet{got07b} 
with $z<0.167$ and $-22.5 < M_r < -19.5$.
The companions/candidates are selected in
the aperture of a 50-kpc radius in which 
each parent galaxy is centrally placed, 
so that they satisfy $-22.5 < M_r < -19.5$.
The redshift criterion of control sample guarantees 
that the sample is volume-limited 
both for the parents and the companions.
We analysed these galaxies as follows.


\begin{table}
\begin{center}
\caption{Number count of true companions $n_{\rm t}$,
estimated number of all companions $n_{\rm et}$ and
expected number of companions per a parent galaxy
$\Phi$,
for the control sample within each defined magnitude.
 }\label{table:Phi_list} {\tabcolsep=5.00mm
\begin{tabular}{cccc}
\hline
\hline
\multicolumn{1}{c}{} &
\multicolumn{1}{c}{$n_{\rm t}$} &
\multicolumn{1}{c}{$n_{\rm et}$} &
\multicolumn{1}{c}{$\Phi$} \\
\hline
{ $-20.5$$\le$$M_r$$<$$-19.5$} & 109  & 276 & 0.0245 \\
{ $-21.5$$\le$$M_r$$<$$-20.5$} & 93  & 235 & 0.0209 \\
{ $-22.5$$\le$$M_r$$<$$-21.5$} & 41 & 104 & 0.00921 \\
 all                           & 243 & 615 & 0.0546 \\
\hline
\end{tabular} 
}
\end{center}
\end{table}

\subsection{Number counts within FOVs of a 50-kpc radius}
\label{section:number_count}

For calculating probability in equation (\ref{eq:p}),
number counts of fore/background galaxies are required.
\citet{yas01} provides the number counts of galaxies using SDSS data.
The result shown in \citet{yas01} is the average
of all observed region.
However, galaxies are not randomly distributed but
tend to cluster.
Therefore, we can expect that
the number count around a galaxy would be higher 
than that of random field by \citet{yas01}.

The required data for our analysis 
are the number counts of the field of view
within a 50-kpc radius in which
each parent galaxy is centrally placed.
The corresponding apparent size of 50-kpc radius 
is changed according to the redshift of the parent galaxy.
Therefore, we construct the data of number counts of
galaxies per area for 5 redshift ranges,
0.03-0.06, 0.06-0.09, 0.09-0.12, 0.12-0.15 and 0.15-0.18.

The $r$-band number count in 50-kpc circular region 
of each redshift range, $\nu_{50} (z,r)$, is calculated by
\begin{equation}
\nu_{50} = \frac{1}{N_{\rm p}} \sum_{j}^{N_{\rm p}} 
 \left(
 \sum_{i}^{n_{{\rm f}_j}} \frac{1}{A_{{\rm fov}_j}}
 \right),
\end{equation}
where $N_{\rm p}$ is the number of parent galaxies
($-22.5 < M_r < -19.5$)
in the corresponding redshift range,
$n_{{\rm f}_j}$ is the number of found galaxies 
in the apparent magnitude range
in the FOV of 50-kpc radius around the parent galaxy $j$,
and $A_{{\rm fov}_j}$ is the area of FOV.
We randomly selected 42,669 parent galaxies
within $0.03 < z <0.18$ and $-22.5 < M_r < -19.5$
from all galaxies in SDSS DR5.
The galaxies within a 50-kpc radius was selected by
executing the {\tt fGetNearbyObjEq()} function 
built in the SDSS SQL service for each parent galaxy.

In total 78,960 galaxies of $13.0 < r < 21.5$
were selected by {\tt fGetNearbyObjEq()}.
The result of $\nu_{50}$ for each redshift range
is shown in Table \ref{table:ncount_list}
and Figure \ref{fig:ncount}.
This figure shows that we cannot ignore the effects of
galaxy correlation in our study.
In section
\ref{section:control_sample} and 
\ref{section:ea_sample},
we use
$\nu_{50} (z,r)$ values in Table \ref{table:ncount_list}
for the number counts in the FOVs
when applying%
\footnote{We did not use any interpolations 
when applying values of Table \ref{table:ncount_list},
since the interpolations did not affect our statistical results.
}
equation (\ref{eq:p}).


%

\subsection{Control sample}
\label{section:control_sample}

After constructing the catalogue of companion candidates
which belongs 11,267 normal galaxies%
\footnote{These galaxies include 15 E+As.
However, excluding these E+As does not affect our 
statistical results.}
within $0.0570 \le z \le 0.0620$,
we divided the candidates into three groups 
according to the absolute magnitude of the $r$-band, $M_r$.
That is, we
set the two thresholds of $M_r$, $-20.5$ and $-21.5$,
for candidates.


Even if the magnitude range of a sample obtained from the SDSS photometric
catalogue satisfies the limit of SDSS spectroscopy ($14.5 < r < 17.7$),
the sample is not complete for the spectroscopic data,
for example because of fibre collision.
Therefore, we counted 
$n_{\rm os}$ (accompanying galaxies with spectroscopic data) and $n_{\rm op}$
(accompanying galaxies with photometric data), and
obtained $n_{\rm os}=409$ and $n_{\rm op}=1035$.
In Table \ref{table:Phi_list}, we showed
$n_{\rm t}$ (number of the true companions), 
$n_{\rm et}$ (the estimated number of all companion galaxies
from equation (\ref{eq:n_et})),
and $\Phi(M_r)$ (the expected number of companions 
per a parent galaxy from equation (\ref{eq:Phi}))
of three absolute magnitude ranges.
We can then calculate the number of parent galaxies for each group.
In this control sample,
the number of $G_{\rm t}$ (galaxies that have true companions), 
$N_{\rm t}$, is 241, and that of $G_{\rm u}$ 
(galaxies that have no true companions but have candidates
without spectroscopy), $N_{\rm u}$, is 550. 
The $P_j$ in equation (\ref{eq:N_e}) is calculated by
equation (\ref{eq:P}). 
And $p_i$, the probability that a candidate is a true companion,
is calculated by equation (\ref{eq:p}).
In equation (\ref{eq:p}), $\Phi_i$ and 
$\phi_i$ are required for each companion candidates.
$\Phi_i$ is obtained from Table \ref{table:Phi_list},
and $\phi_i$ is derived as
\begin{equation}
\label{eq:phi}
 \phi_i =  \nu_{50}(z,r) \cdot A_{{\rm fov}_i},
\end{equation}
where $A_{{\rm fov}_i}$ is the area corresponding
to the region of a 50-kpc radius and
$\nu_{50}(z,r)$ is the number count of galaxies
in Table \ref{table:ncount_list}.
We use absolute magnitude of the candidate for $\Phi_i$, 
and use apparent magnitude of the candidate for $\phi_i$.

We obtained $N_{\rm e} = 336$ and $N_{\rm et} = 577$.
This indicates that 5.12\% ($=$$577/11267$) of normal galaxies have
companion galaxies within a 50-kpc radius.
The numbers are shown in Table \ref{table:result_list}.

\begin{table}
\begin{center}
\caption{Summary of the numbers in this study.
$N_{\rm M}$ is the number of parent galaxies, and
$N_{\rm et}$ is the estimated number of parent galaxies with
true companions.  See the text for other notations.
 }\label{table:result_list} {\tabcolsep=1.20mm
\begin{tabular}{lcccccccc}
\hline
\hline
\multicolumn{1}{c}{} &
\multicolumn{1}{c}{$N_{\rm M}$} &
\multicolumn{1}{c}{$n_{\rm op}$} &
\multicolumn{1}{c}{$n_{\rm os}$} &
\multicolumn{1}{c}{$n_{\rm t}$} &
\multicolumn{1}{c}{$N_{\rm t}$} &
\multicolumn{1}{c}{$N_{\rm u}$} &
\multicolumn{1}{c}{$N_{\rm e}$} &
\multicolumn{1}{c}{$N_{\rm et}$} \\
\hline
Control sample & {\bf 11267} & 1035 & 409 & 243 & 241 & 550 & 336 & {\bf 577} \\
E+A sample     & {\bf 660}   & 119 & 41 & 34 & 34 & 63 & 18.0 & {\bf 52.0} \\
\hline
\end{tabular} 
}
\end{center}
\end{table}

\begin{figure*}
 \begin{center}
 \tabcolsep=0.45mm
  \includegraphics[scale=0.745]{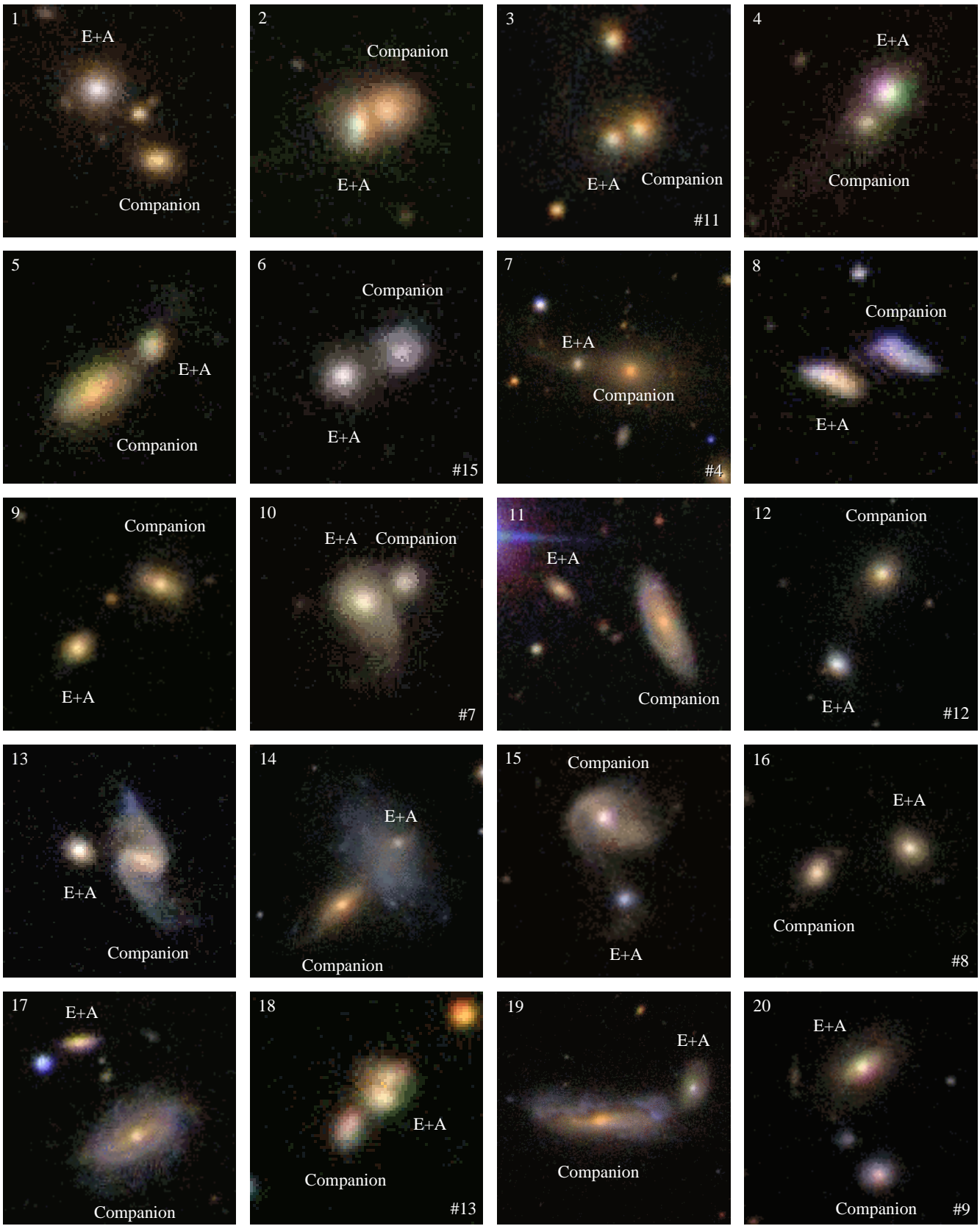}
 \end{center}
 \caption{SDSS $g$,$r$,$i$-composite images of E+A and companion pairs taken from SDSS CAS.
 This figure includes
 pair ID no.1({\it top left}), ... , no.4({\it top right}), ..., 
 no.17({\it bottom left}), ... and no.20({\it bottom right}).
 The inlaid number at the bottom right in each image
 is the target ID of our KPNO observation.
 }\label{fig:images}
\end{figure*}

\begin{figure*}
 \begin{center}
 \tabcolsep=0.45mm
  \includegraphics[scale=0.745]{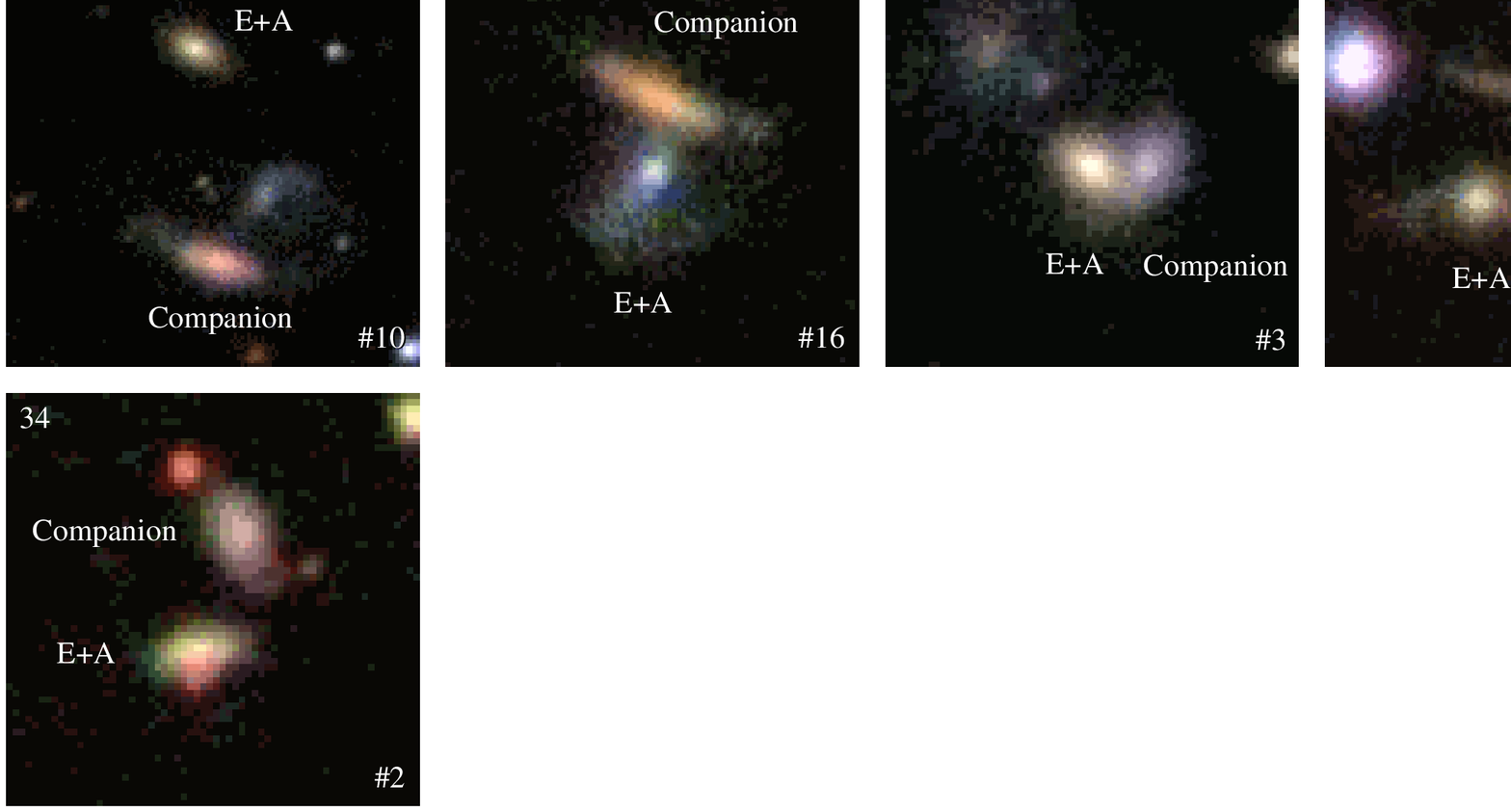}
 \end{center}
 \contcaption{Pair ID
 No.21({\it top left}), ... , No.24({\it top right}), ... and
 No.34({\it bottom left}).
 }
\end{figure*}

\subsection{E+A sample}
\label{section:ea_sample}


Following the description in section \ref{section:samples},
we constructed the companion candidate catalogue for E+A galaxies.
We found $n_{\rm op} = 119$ of
photometrically observed companion candidates,
and $19$ of the candidates were spectroscopically observed in the SDSS.
Our KPNO observations appended $19$ redshifts of companion candidates, and
we found 3 redshifts of candidates from the NED search service.
The number of spectroscopically observed candidates then became
$n_{\rm os} = 41$.
We found 14 true companions from SDSS and
17 true companions from our KPNO observation,
and we found 3 true companions from the NED search service.
That is, 7 objects turned out to be
fore/background galaxies.
Therefore, the number of true companion galaxies $n_{\rm t}$ is 34
in the $g_{\rm os}$.
As a result, $N_{\rm t} = 34$ of E+A galaxies have true companions
and $N_{\rm u} = 63$ of E+As have 
candidates with no spectroscopy.
We present the 34 E+A and companion systems in Table \ref{table:all_list}
and Figure \ref{fig:images}.

We then estimate the number of E+As with companions
($N_{\rm et}$) in the 63 galaxies.
Since the redshift range of E+A sample is large,
the E+A sample does not satisfy the volume-limited condition,
and we cannot estimate the expected numbers of companions $\Phi(M_r)$.
Therefore we used the $\Phi(M_r)$ of the control sample 
(Table \ref{table:Phi_list}) for the estimation of number of 
$G_{\rm u}$ E+A galaxies,
following the null hypothesis that the probability of existence of
companion is the same for E+A and normal galaxies.
%
However, if the fraction of parent galaxies with companions
in the E+A sample is really larger than that in the control sample,
$N_{\rm et} / N_{\rm M}$ of the E+A sample
and that of the control sample will show a significant difference
even adopting the hypothesis. 
As described in section \ref{section:control_sample}, 
we estimate the number of E+As with companions ($N_{\rm et}$).
Equation (\ref{eq:P}) -- (\ref{eq:N_et}) resulted that
the estimated $N_{\rm e}$ and $N_{\rm et}$ of E+A sample were 
18.0 
and 52.0, respectively. 
These results are summarised in Table \ref{table:result_list}.

The result indicates that at least 7.88\% ($=$$52.0/660$) 
of E+A galaxies have companion galaxies.
In section \ref{section:control_sample}, 
we found 5.12\% ($=$$577/11267$) of normal galaxies have at least one companion, i.e., 
E+A galaxies have 54\% higher probability to do that.
 How statistically significant is this result?
We applied a two-sample test for equality of proportions 
with continuity correction to answer the question.
We used the {\tt prop.test()} function of 
R (http://www.r-project.org/) version 2.0.1.
We set the parameter as
 {\tt prop.test(c(52.0,577), c(660,11267))}, 
 whereupon we achieved a 99.7\% level of significance
 that the existing probabilities of companion galaxies are statistically different between the E+A (7.88\%) and control samples (5.12\%).
 
 Although it was pointed out previously that E+A galaxies might have more companion galaxies than normal galaxies, 
 this work is the first to show the result 
based on spectroscopic data, with statistical meaningful significance.

\section{Discussion and Summary}
\label{section:discussion}

We constructed the first catalogue of
E+A with companion galaxies based on spectroscopic data.
The catalogue provides a basis for discussing the
statistical analysis of E+A and companion systems.
In section \ref{section:analysis_and_results}, we statistically showed
that E+A galaxies have by 54\% more companion galaxies than the control sample
of normal (average) galaxies within 50 kpc.
 This result is a major step forward on the subject in that the analysis is based on the {\boldmath spectroscopic} redshift survey of companion candidates;
 Although previous studies have suggested that E+A galaxies have more
 companions, some were based on mere a morphological impression of E+A
 galaxies \citep{oeg91,cou94,cou98,dre94,oem97,yan04,yam05}, while
 others relied on the statistical analysis of imaging data
 \citep{got05}, which suffered from a large uncertainty from
 fore/background galaxies. 
 Our findings are much more reliable in that the analysis is based on
 spectroscopic data; that is, the 34 companion galaxies found in this
 work truly are in physically close proximity where they can
 dynamically interact with the central E+A galaxy.
 Therefore, our finding greatly strengthens the physical
 interpretation of the result; physical interaction/merger with 
 companion galaxies is likely the origin of E+A galaxies. 

Previous observational studies, for example, morphologies, radial
photometric/spectroscopic analysis
\citep{bar01,yan04,yam05}
and numerical simulations \citep{bek01,bek05} become much more realistic with
our redshift identifications of companion galaxies.
Taken together, little doubt exists that the
merger/interaction is an important aspect in understanding the evolution
of E+A galaxies. 

Although we showed that E+A galaxies have 54\% more companions than normal galaxies,
 our analysis also showed that less than 10\% of E+A galaxies have
companion galaxies.  This result implies
that the origin of a large fraction of E+A galaxies may not be a flyby interaction
but a galaxy--galaxy merger
(i.e., a parent galaxy engulfed its companion(s)).

Several caveats must be kept in mind.
A merger/interaction can enhance star formation in galaxies.
\citet{lam03} examined 1,258 galaxy pairs in the 100k public release of
the 2dF galaxy survey and found that star formation in galaxy pairs is
significantly enhanced over that of isolated galaxies for separations
less than 36 kpc and velocity differences less than 100 km s$^{-1}$.
In addition,
\citet{nik04} also investigated
12,492 galaxy pairs at 
projected separations of less than 300 kpc using the SDSS DR1,
and reported that the mean specific star formation rate is 
significantly enhanced for projected separations of less than 30 kpc.
E+A galaxies in our catalogue may evolve from these galaxies
with violent star formation.
Therefore, not every merging/interacting galaxy is an E+A galaxy.
The fraction of E+A galaxies in the nearby Universe is too small
($\sim$0.02\%; Goto 2005) for every merging/interacting galaxy to go through, even
if the short timescale of the E+A phase ($\sim$1 Gyr) is considered. In
our spectroscopic survey of E+A companion galaxies, some (true)
companion galaxies are star-forming galaxies and other companion
galaxies are passive (elliptical) galaxies. 
We found only one pair in this study in which a companion of an E+A
galaxy is also an E+A galaxy. 
Therefore, another condition must exist for a merging/interacting
galaxy to become an E+A. 

 Our work is based on local E+A galaxies.
 However, in high-redshift cluster environments, the situation is
 different; E+A galaxies are much more numerous. 
 Pioneering work was done by
 \citet{dre99,pog99,dre04}, who found that E+A galaxies 
 ($3{\rm \AA} < {\rm H}\delta~{\rm EW}$ and undetectable emission in
 ${\rm [OII]}$) are significantly more
 common in 10 clusters at $0.37<z<0.56$ than in the field ($21\pm2$\%
 compared to $6\pm3$\%).
 Later, \citet{tra03}  found 7--13\% of E+A galaxies in three
 high-redshift clusters at $z=0.33,0.58$ and $0.83$, claiming that
 $>30$\% of E+S0 members may have undergone the E+A phase if the effects
 of E+A downsizing and increasing E+A fraction as a function of redshift
 are considered (their selection criteria was 
 $4{\rm \AA} < \frac{{\rm H}\delta {\rm EW}+{\rm H}\gamma {\rm EW}}{2}$
 and $-5{\rm \AA} < {\rm [OII]} {\rm EW}$). 
 In their search for field E+A
 galaxies amongst 800 spectra, \citet{tra04} measured the E+A fraction
 at $0.3 < z < 1$ to be $2.7$\% $\pm$ $1.1$\%, a value lower than that in
 galaxy clusters at comparable redshifts.
 When we refer to E+A galaxies, although we are looking at the same
 evolutionary stage of galaxies, it is important to keep in mind that
 the origin of E+A galaxies might be heterogeneous. For example,
 high-redshift cluster E+A galaxies may have a different physical origin
 from that of local field E+A galaxies.

 We should also keep in mind that the magnitude range of our sample is
 limited to relatively bright magnitude ($-22.5 < M_r < -19.5$). 
 It is expected that the fraction of E+A galaxies with a companion
 galaxy increases as we expand our spectroscopic survey of companion
 galaxies to fainter magnitude at $-19.5 < M_r$. It is also of
 importance to investigate a wide range of absolute magnitudes to reveal
 the luminosity dependence of the E+A phase. Magnitudes of the merger
 (major/minor) may have some effects in creating E+A galaxies. The time is
 right to consider deeper spectroscopic surveys of E+A companion galaxies
 with 4- to 8-m class telescopes.

Thus, our next goal is to clarify the evolution of E+A systems. For that,
more detailed studies \citep[e.g.][]{yag06a,yag06b,got07a} are needed.
It is critical to investigate the spectroscopic properties of E+A
companion galaxies and their dependence on the E+A properties. 
Revealing the effect of the local environment on E+A galaxies is also
important future work.

\section*{Acknowledgements}

We thank Diane Harmer, Daryl Willmarth, Judy Prosser and many KPNO
staff members for much guidance and help in our preparations and
observations.
We are grateful to Dr. Hiroyasu Ando (NAOJ),
Dr. Kunio Noguchi (NAOJ),
Dr. Hiroshi Murakami (ISAS/JAXA),  
Dr. Takao Nakagawa (ISAS/JAXA) and
M. S. Vijaya Kumar (TMU) for their support.
We thank the anonymous referee for many insightful comments,
which improved the paper significantly.

The research was supported by the Hayakawa Fund from
the Astronomical Society of Japan.


Funding for the creation and distribution of the SDSS Archive
has been provided by the Alfred P. Sloan Foundation, the Participating
Institutions, the National Aeronautics and Space Administration, the
National Science Foundation, the U.S. Department of Energy, the Japanese
Monbukagakusho, and the Max Planck Society. The SDSS Web site is
http://www.sdss.org/.

SDSS is managed by the Astrophysical Research Consortium (ARC) for
the participating institutions, which are the
University of Chicago, Fermilab, the Institute for Advanced Study, the
Japan Participation Group, Johns Hopkins University, Los Alamos
National Laboratory, the Max Planck Institute for Astronomy (MPIA), the
Max Planck Institute for Astrophysics (MPA), New Mexico State
University, University of Pittsburgh, Princeton University, the United
States Naval Observatory and the University of Washington.

This research made use of the NASA/IPAC Extragalactic Database (NED)
operated by the Jet Propulsion Laboratory, California Institute
of Technology, under contract with the National Aeronautics and Space
Administration.

We thank Linux, XFree86, IRAF and other UNIX-related communities
for the development of various useful software.
This research made use of the Plamo Linux.

\section*{Appendix1:~~Spectra of faint targets observed with the KPNO 2.1-m telescope}

\begin{figure}
 \includegraphics[scale=0.32]{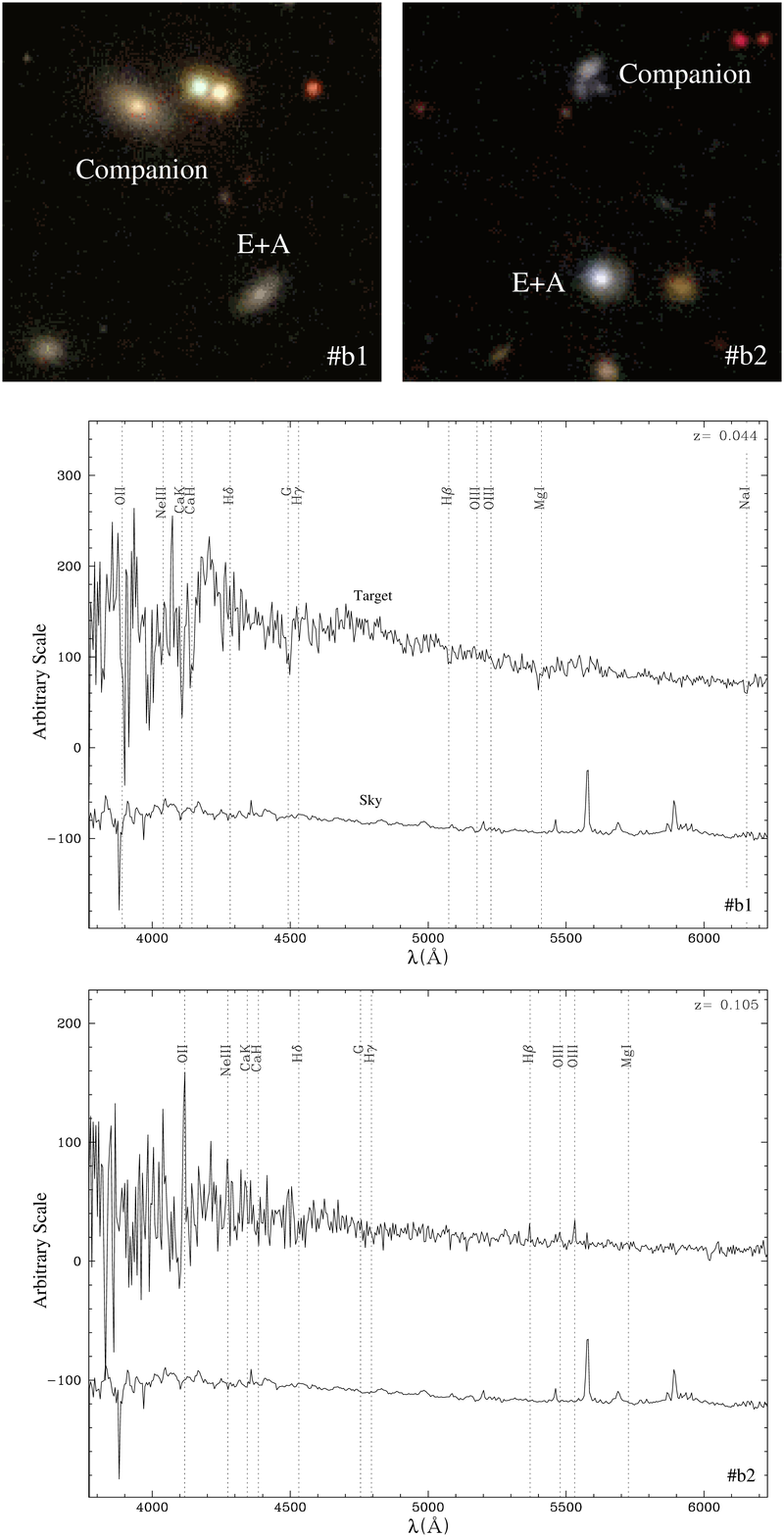}
 \caption{SDSS $g$,$r$,$i$-composite images and spectra of two new E+A's
 companions observed as backup targets
 with the KPNO 2.1-m telescope applying a 20\AA~binning. 
 A sky spectrum in the observed data is shown at the bottom of each
 panel.
 Features to identify the redshift of candidates
 are ${\rm CaK}$ and ${\rm CaH}$ absorption lines (\#b1) and
 ${\rm [OII]}$, ${\rm H}\beta$ and ${\rm [OIII]}$ emission lines (\#b2), 
 respectively.
 }\label{fig:spectra_backup}
\end{figure}

We actually observed two additional companion candidates as backup targets
using the GoldCam at the KPNO.
We show their spectra in Figure \ref{fig:spectra_backup}
and the target list is presented in
Table \ref{table:comp_list_backup}.
These backup targets do not have magnitudes that satisfy $M_r < -19.5$
for both the parent galaxy and companion candidates.
In addition, the
projection distance of target b2
is larger than 50 kpc.
However, we found that these two candidates are also true companion
galaxies.
This result may contribute to 
future studies.

\begin{table*}
\begin{center}
\caption{List of backup targets with
 successful observation with the KPNO 2.1-m telescope.
 }\label{table:comp_list_backup} {\tabcolsep=0.70mm
\begin{tabular}{cccccccccccccccc}
\hline
\hline
\multicolumn{1}{c}{~} &
\multicolumn{1}{c}{~} &
\multicolumn{1}{c}{Exposure} &
\multicolumn{6}{c}{E+A galaxies} &
\multicolumn{5}{c}{Companion galaxies} &
\multicolumn{1}{c}{~} &
\multicolumn{1}{c}{\scriptsize projection} \\
\multicolumn{1}{c}{Target} &
\multicolumn{1}{c}{Observation} &
\multicolumn{1}{c}{time} &
\multicolumn{1}{c}{R.A.} &
\multicolumn{1}{c}{Dec.} &
\multicolumn{1}{c}{$r$} &
\multicolumn{1}{c}{$M_r$} &
\multicolumn{1}{c}{$z$} &
\multicolumn{1}{c}{${\rm H}\delta~{\rm EW}$} &
\multicolumn{1}{c}{R.A.} &
\multicolumn{1}{c}{Dec.} &
\multicolumn{1}{c}{$r$} &
\multicolumn{1}{c}{$M_r$} &
\multicolumn{1}{c}{$z$} &
\multicolumn{1}{c}{$|\Delta z|$} &
\multicolumn{1}{c}{\scriptsize distance(kpc)} \\
\hline
 \#b1 & Sep.22,2005 & 30min.$\times$2 & 
 01:15:37.59 & +00:05:34.0 & 17.60 & -18.80 & 0.044 & 4.50 &
 01:15:39.20 & +00:06:11.0 & 16.11 & -20.29 & 0.044 &
 0.000 & 37.42 \\
 \#b2 & Sep.24,2005 & 30min.$\times$4 &
 17:05:57.44 & +63:00:55.7 & 16.66 & -21.75 & 0.105 & 7.00 &
 17:05:57.76 & +63:01:33.2 & 18.50 & -19.90 & 0.105 &
 0.000 & 71.88 \\
\hline
\end{tabular} 
}
\end{center}
\end{table*}

\section*{Appendix2:~~Spectra of fore/background galaxies observed at KPNO 2.1-m telescope}

\begin{figure}
 \includegraphics[scale=0.32]{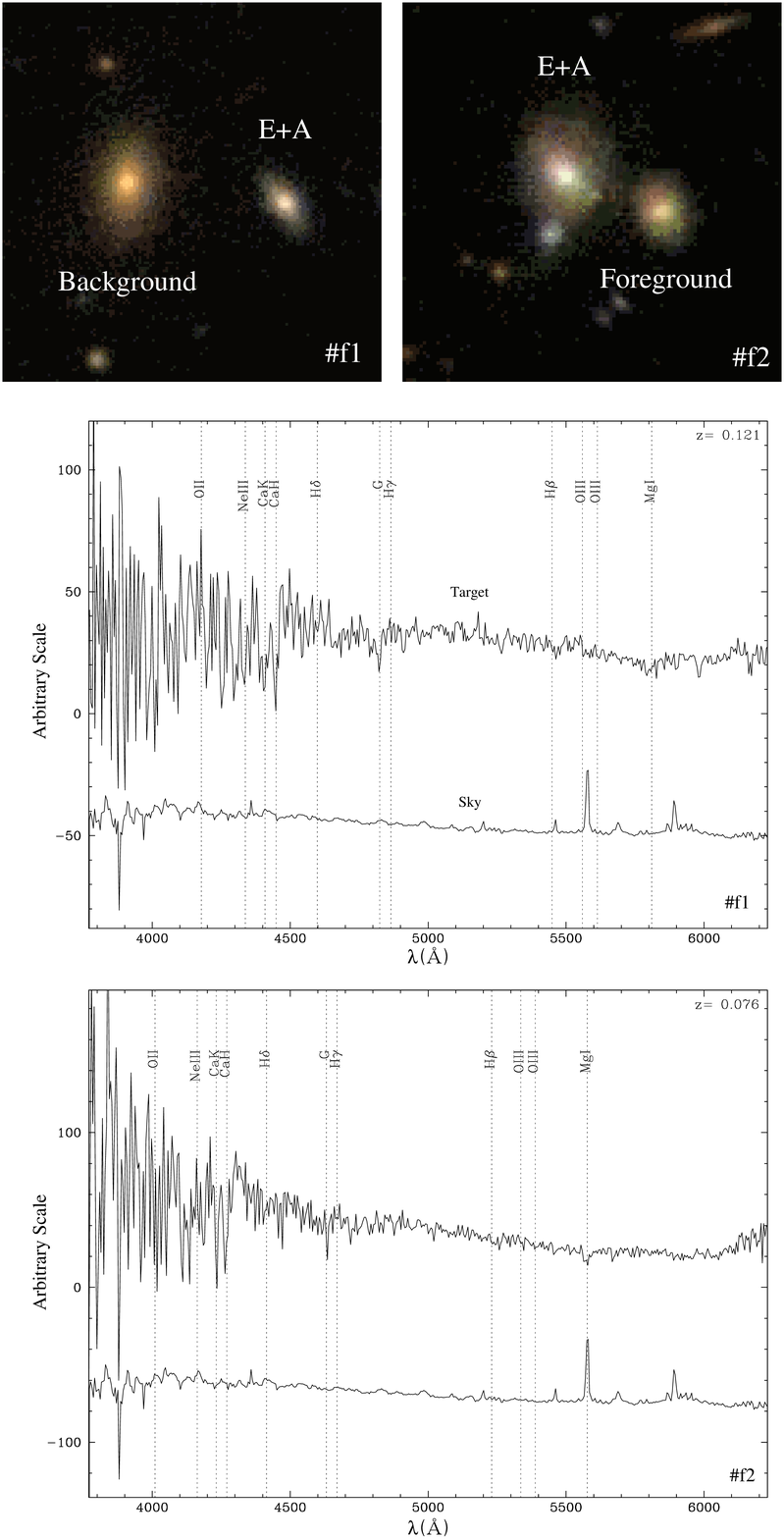}
 \caption{SDSS $g$,$r$,$i$-composite images and spectra of two
 fore/background galaxies taken 
 with the KPNO 2.1-m telescope applying a 20\AA~binning. 
 A sky spectrum in the observed data is shown at the bottom of each
 panel.
 Features to identify the redshift of candidates
 are ${\rm CaK}$ and ${\rm CaH}$ absorption lines.
 }\label{fig:spectra_fake}
\end{figure}

We identified two fore/background galaxies around E+A galaxies
by our KPNO observations.
Results are shown in Figure \ref{fig:spectra_fake} and 
Table \ref{table:fake_list}.

\begin{table*}
\begin{center}
\caption{List of fore/background galaxies with
 successful observation with the KPNO 2.1-m telescope.
 }\label{table:fake_list} {\tabcolsep=0.70mm
\begin{tabular}{cccccccccccccc}
\hline
\hline
\multicolumn{1}{c}{~} &
\multicolumn{1}{c}{~} &
\multicolumn{1}{c}{Exposure} &
\multicolumn{6}{c}{E+A galaxies} &
\multicolumn{5}{c}{Fore/Background galaxies}\\
\multicolumn{1}{c}{Target} &
\multicolumn{1}{c}{Observation} &
\multicolumn{1}{c}{time} &
\multicolumn{1}{c}{R.A.} &
\multicolumn{1}{c}{Dec.} &
\multicolumn{1}{c}{$r$} &
\multicolumn{1}{c}{$M_r$} &
\multicolumn{1}{c}{$z$} &
\multicolumn{1}{c}{${\rm H}\delta~{\rm EW}$} &
\multicolumn{1}{c}{R.A.} &
\multicolumn{1}{c}{Dec.} &
\multicolumn{1}{c}{$r$} &
\multicolumn{1}{c}{$M_r$} &
\multicolumn{1}{c}{$z$} \\
\hline
 \#f1 & May.19,2007 & 20min.$\times$3 & 
 11:22:33.75 & +31:35:06.2 & 17.26 & -21.34 & 0.114 & 4.54 &
 11:22:35.57 & +31:35:09.2 & 16.44 & -22.29 & 0.121 \\
 \#f2 & May.20,2007 & 20min.$\times$9 &
 12:08:11.28 & +40:21:51.2 & 16.42 & -22.15 & 0.113 & 5.83 &
 12:08:10.28 & +40:21:47.0 & 17.93 & -19.73 & 0.076 \\
\hline
\end{tabular} 
}
\end{center}
\end{table*}

\label{lastpage}


\begin{thebibliography}{DUM}

 \bibitem[Abadi, Moore \& Bower(1999)]{aba99}
	Abadi, M. G., Moore, B., \& Bower, R. G.
	1999, \mnras, 308, 947

 \bibitem[Abazajian et al.(2003)]{aba03}
	Abazajian, K., et al. 2003, \aj, 126, 2081

 \bibitem[Abazajian et al.(2004)]{aba04}
	Abazajian, K., et al. 2004, \aj, 128, 502

 \bibitem[Abazajian et al.(2005)]{aba05}
	Abazajian, K., et al. 2005, \aj, 129, 1755

 \bibitem[Abraham et al.(1996)]{abr96}
	Abraham, R. G., et al. 1996, \apj, 471, 694

 \bibitem[Adelman-McCarthy et al.(2006)]{ade06}
	 Adelman-McCarthy, J., et al. 2006, \apjs, 162, 38

 \bibitem[Adelman-McCarthy et al.(2007)]{ade07}
	 Adelman-McCarthy, J., et al. 2007, \apjs, 172, 634

 \bibitem[Buyle et al.(2006)]{buy06}
	Buyle, P., Michielsen, D., De Rijcke, S., Pisano, D. J.,
	Dejonghe, H., and Freeman, K.
	2006, \apj, 649, 163

 \bibitem[Bartholomew, Rose \& Gaba(2001)]{bar01}
	Bartholomew, L. J., Rose, J. A., \& Gaba, A. E.
	2001, \aj, 122, 2913


 \bibitem[Barger et al.(1996)]{bar96}
	Barger, A. J., Aragon-Salamanca, A., Ellis, R. S.,
	Couch, W. J., Smail, I., \&
	Sharples, R. M.
	1996, \mnras, 279, 1





 \bibitem[Bekki, Shioya \& Couch(2001)]{bek01}
	Bekki, K., Shioya, Y., \& Couch, W. J.
	2001, \apj, 547, L17

 \bibitem[Bekki et al.(2005)]{bek05}
	Bekki, K., Couch, W. J., Shioya, Y., \&  Vazdekis, A.
	2005, \mnras, 359, 949

 \bibitem[Belloni et al.(1995)]{bel95}
	Belloni, P., Bruzual, A. G., Thimm, G. J., \&
	Roser, H.-J.
	1995, A\&A, 297, 61

 \bibitem[Bennet et al.(2003)]{ben03}
	Bennett, C. L., et al. 
	2003, \apjs, 148, 1

 \bibitem[Blake et al.(2004)]{bla04}
	Blake, C., et al. 2004
	\mnras, 355, 713

 \bibitem[Blanton et al.(2003)]{bla03}
	Blanton, M. R., et al. 2003,  
	\aj, 125, 2348

 \bibitem[Broadhurst, Ellis \& Shanks(1988)]{bro88}
	Broadhurst, T. J., Ellis, R. S., \& Shanks, T.
	1988, \mnras, 235, 827

 \bibitem[Butcher \& Oemler(1978)]{but78}
	Butcher, H., \& Oemler, A.
	1978, \apj, 226, 559 

 \bibitem[Caldwell et al.(1993)]{cal93}
	Caldwell, N., Rose, J. A., Sharples, R. M.,
	Elllis, R. S., \& Bower, R. G.
	1993, \aj, 106, 473


 \bibitem[Caldwell \& Rose(1997)]{cal97}
	Caldwell, N., \& Rose, J. A.
	1997, \aj, 113, 492

 \bibitem[Castander et al.(2001)]{cas01}
	Castander, F. J. et al.
	2001, \aj, 121, 2331

 \bibitem[Chang et al.(2001)]{cha01}
	Chang, T., van Gorkom, J. H., Zabludoff, A. I.,
	Zaritsky, D., \& Mihos, J. C.
	2001, \aj, 121, 1965

 \bibitem[Condon(1992)]{con92}
	Condon, J. J.
	1992, ARA\&A, 30, 575

 \bibitem[Couch \& Sharples(1987)]{cou87}
	Couch, W. J., \& Sharples, R. M.
	1987, \mnras, 229, 423

 \bibitem[Couch et al.(1994)]{cou94}
	Couch, W. J., Ellis, R. S., Sharples, R. M., \& Smail, I.
	1994, \apj, 430, 121

 \bibitem[Couch et al.(1998)]{cou98}
	Couch, W. J., Barger, A. J., Smail, I., Ellis, R. S., \&
	Sharples, R. M.
	1998, \apj, 497, 188



 \bibitem[Dressler(1980)]{dre80}
	Dressler, A.
	1980, \apj, 236, 351 

 \bibitem[Dressler \& Gunn(1983)]{dre83}
	Dressler, A., \& Gunn, J. E.
	1983, \apj, 270, 7 

 \bibitem[Dressler \& Gunn(1992)]{dre92}
	Dressler, A., \& Gunn, J. E.
	1992, \apjs, 78, 1 

 \bibitem[Dressler et al.(1994)]{dre94}
	Dressler, A., Oemler, A. J., Sparks, W. B., \& Lucas, R. A.
	1994, \apj, 435, L23

 \bibitem[Dressler et al.(1999)]{dre99}
	Dressler, A., Smail, I., Poggianti, B. M., Butcher, H.,
	Couch, W. J., Ellis, R. S., \& Oemler, A. J.
	1999, \apjs, 122, 51

 \bibitem[Dressler et al.(2004)]{dre04}
	Dressler, A., Oemler, A. Jr., Poggianti, B. M., 
        Smail, I., Trager, S., Shectman, S. A., Couch, W. J., 
        Ellis, R. S.
	2004, \apj, 617, 867

 \bibitem[Ellingson et al.(2001)]{ell01} 
	Ellingson, E., Lin, H., Yee, H. K. C., \&
	Carlberg, R. G.
	2001, \apj, 547, 609 

 \bibitem[Fabricant, McClintock \& Bautz(1991)]{fab91}
	Fabricant, D. G., McClintock, J. E., \& Bautz, M. W.
	1991, \apj, 381, 33

 \bibitem[Farouki \& Shapiro(1980)]{far80}
	Farouki, R., \& Shapiro, S. L.
	1980, \apj, 241, 928

 \bibitem[Fasano et al.(2000)]{fas00}
	Fasano, G., Poggianti, B. M.,
	Couch, W. J., Bettoni, D.,
	Kj{\ae}rgaard, P., \& Moles, M.
	2000, \apj, 542, 673


 \bibitem[Fisher et al.(1998)]{fis98}
	Fisher, D., Fabricant, D., Franx, M., \&
	van Dokkum, P.
	1998, \apj, 498, 195

 \bibitem[Franx(1993)]{fra93}
	Franx, M. 1993, \apj, 407, L5

 \bibitem[Fujita \& Nagashima(1999)]{fuj99}
	Fujita, Y., \& Nagashima, M. 1999, \apj, 516, 619

 \bibitem[Fujita(2004)]{fuj04}
	Fujita, Y. 2004, \pasj, 56, 29

 \bibitem[Fujita \& Goto(2004)]{fg04}
	Fujita, Y., \& Goto, T.
	2004, \pasj, 56, 621

 \bibitem[Fukugita et al.(1996)]{fuk96}  
        Fukugita, M., Ichikawa, T., Gunn, J. E., Doi, M.,
        Shimasaku, K., \& Schneider, D. P. 
        1996, \aj, 111, 1748

 \bibitem[Goto(2003)]{got03}
	Goto, T., 2003, PhD Dissertation, 
	University of Tokyo, astro-ph/0310196

 \bibitem[Goto et al.(2003a)]{got03a}
	Goto, T. et al. 2003a, \pasj, 55, 739

 \bibitem[Goto et al.(2003b)]{got03b}
	Goto, T. et al. 2003b, \pasj, 55, 771


 \bibitem[Goto et al.(2003c)]{got03c}
	Goto, T., Yamauchi, C., Fujita, Y., Okamura, S.,
	Sekiguchi, M., Smail, I., Bernardi, M., \&
	Gomez, P. L.
	2003c, \mnras, 346, 601

 \bibitem[Goto(2004)]{got04a}
	Goto, T.
	2004, \aap, 427, 125

 \bibitem[Goto(2005)]{got05}
 	Goto, T.
 	2005, \mnras, 357, 937

 \bibitem[Goto(2007a)]{got07a}
 	Goto, T.
 	2007, \mnras, 377, 1222

 \bibitem[Goto(2007b)]{got07b}
 	Goto, T.
 	2007, \mnras, 381, 187

 \bibitem[Gunn \& Gott(1972)]{gun72}
	Gunn, J. E., \& Gott, J. R. I.
	1972, \apj, 176, 1

 \bibitem[Gunn et al.(1998)]{gun98} 
	Gunn, J. E., et al. 1998, \aj, 116, 3040

 \bibitem[Hogg et al.(2001)]{hog01} 
        Hogg, D. W., Schlegel, D. J., \&
        Finkbeiner, D. P., 
        \& Gunn, J. E. 2001, \aj, 122, 2129

 \bibitem[Hogg et al.(2006)]{hog06}
	Hogg, D. W.,
	Masjedi, M.,
	Berlind, A. A.,
	Blanton, M. R.,
	Quintero, A. D.,
	\& Brinkmann, J. 2006, \apj, 650, 763

 \bibitem[Hopkins et al.(2003)]{hop03}
	Hopkins, A. M., et al. 
	2003, \apj, 599, 971



 \bibitem[Kauffmann et al.(2003)]{kau03}
        Kauffmann, G., et al.
        2003, \mnras, 341, 33

 \bibitem[Kennicutt(1998)]{ken98}
	Kennicutt, R. C. 
	1998, ARA\&A, 36 189

 \bibitem[Kent(1981)]{ken81}
	Kent, S. M.
	1981, \apj, 245, 805

 \bibitem[Kodama \& Bower(2001)]{kod01}
	Kodama, T. \& Bower, R. G.
	2001, \mnras, 321, 18 

 \bibitem[Komatsu et al.(2008)]{kom08}
	Komatsu, E., et al.
	2008, \apjs, submitted


 \bibitem[Lambas(2003)]{lam03}
	Lambas, D. G., Tissera, P. B., Sol Alonso, M.,
	\& Coldwell, G.
	2003, \mnras, 346, 1189

 \bibitem[Lavery \& Henry(1986)]{lav86}
	Lavery, R. J. \& Henry, J. P.
	1986, \apj, 304, L5

 \bibitem[Lavery \& Henry(1988)]{lav88}
	Lavery, R. J. \& Henry, J. P.
	1988, \apj, 330, 596

 \bibitem[Liu \& Kennicutt(1995a)]{liu95a}
	Liu, C. T., \& Kennicutt, R. C.
	1995a, \apjs, 100, 325

 \bibitem[Liu \& Kennicutt(1995b)]{liu95b}
	Liu, C. T., \& Kennicutt, R. C.
	1995b, \apj, 450, 547


 \bibitem[MacLaren, Ellis \& Couch(1988)]{mac88}
	MacLarn, I., Ellis, R. S., \& Couch, W. J.
	1988, \mnras, 230, 249

 
 \bibitem[Margoniner et al.(2001)]{mar01} 
	Margoniner V. E., de Carvalho R. R., Gal R. R., Djorgovski S. G.
	2001, \apj, 548, L143


 \bibitem[Miller \& Owen(2001)]{mil01}
	Miller, N. A., \& Owen, F. N.
	2001, \apj, 554, L25




 \bibitem[Morris et al.(1998)]{mor98}
	Morris, S. J., Hutchings, J. B., Carlberg, R. G.,
	Yee, H. K. C., Ellingson, E., Balogh, M. L.,
	Abraham, R. G., \& Smecker-Hane, T. A.
	1998, \apj, 507, 84

 \bibitem[Newberry, Boroson \& Kirshner(1990)]{new90}
	Newberry, M. V., Boroson, T. A., \& Kirshner, R. P.
	1990, \apj, 350, 585

 \bibitem[Nikolic, Cullen \& Alexander(2004)]{nik04}
	Nikolic, B., Cullen, H., \& Alexander, P.
	2004, \mnras, 355, 874

 \bibitem[Norton et al.(2001)]{nor01}
	Norton, S. A., Gebhardt, K., Zabludoff, A. I., \&
	Zaritsky, D.
	2001, \apj, 557, 150

 \bibitem[Oegerle, Hill \& Hoessel(1991)]{oeg91}
	Oegerle, W. R., Hill, J. M., \& Hoessel, J. G.
	1991, \apj, 381, L9

 \bibitem[Oemler, Dressler \& Butcher(1997)]{oem97}
	Oemler, A. J., Dressler, A., \& Butcher, H. R.
	1997, \apj, 474, 561


 \bibitem[Pier et al.(2003)]{pie03}  
        Pier, J. R., Munn, J. A., Hindsley, R. B.,
        Hennessy, G. S., Kent, S. M.,
        Lupton, R. H., \& Ivezic, Z.
        2003, \aj, 125, 1559

 \bibitem[Poggianti et al.(1999)]{pog99}
	Poggianti, B. M., Smail, I., Dressler, Alan., Couch, W. J., 
        Barger, A. J., Butcher, H., Ellis, R. S., \& Oemler, A. Jr.
	1999, \apj, 518, 576

 \bibitem[Poggianti \& Wu(2000)]{pog00}
	Poggianti, B. M., \& Wu, H.
	2000, \apj, 529, 157

 \bibitem[Popesso et al.(2007)]{pop07}
	Popesso, P.,
	Biviano, A.,
	Romaniello, M., \& 
	B\"ohringer, H. 2007, \aap, 461, 411

 \bibitem[Postman \& Geller(1984)]{pos84}
	Postman M., Geller M. J.
	1984, \apj, 281, 95
 
 \bibitem[Postman et al.(2005)]{pos05}
	Postman M., et al.
	2005, \apj, 623, 721

 \bibitem[Quilis, Moore \& Bower(2000)]{qui00}
	Quilis, V., Moore, B., \& Bower, R.
	2000, Sci, 288, 1617

 \bibitem[Quintero et al.(2004)]{qui04}
	Quintero, A. D., et al. 2004, \apj, 602, 190

 \bibitem[Rakos \& Schombert(1995)]{rak95}
	Rakos, K. D., \& Schombert, J. M.
	1995, \apj, 439, 47

 \bibitem[Rose et al.(2001)]{ros01}
	Rose, J. A., Gaba, A. E., Caldwell, N., \&
	Chaboyer, B. 2001, \aj, 121, 793



 \bibitem[Schweizer(1982)]{sch82}
	Schweizer, F. 1982, \apj, 252, 455

 \bibitem[Schweizer(1996)]{sch96}
	Schweizer, F. 1996, \aj, 111, 109

 \bibitem[Sharples et al.(1985)]{sha85}
	Sharples, R. M., Ellis, R. S., Couch, W. J., \&
	Gray, P. M.
	1985, \mnras, 212, 687



 \bibitem[Smail et al.(1999)]{sma99}
	Smail, I., Morrison, G., Gray, M. E., Owen, F. N.,
	Ivison, R. J., Kneib, J.-P., \& Ellis, R. S.
	1999, \apj, 525, 609

 \bibitem[Smith et al.(2002)]{smi02}  
        Smith, J. A., et al. 2002, \aj, 123, 2121

 \bibitem[Smith et al.(2005)]{smi05}
	Smith, G. P.,
	Treu, T., Ellis, R. S., Moran, S. M., \& Dressler, A.
	2005, \apj, 620, 78S

 \bibitem[Spitzer \& Baade(1951)]{spi51}
	Spitzer, L. J., \& Baade, W.
	1951, \apj, 113, 413

 \bibitem[Stoughton et al.(2002)]{sto02} 
        Stoughton, C., et al. 2002, \aj, 123, 485


 \bibitem[Strauss et al.(2002)]{str02}  
	Strauss, M. A. et al. 2002, \aj, 124, 1810

 \bibitem[Tanaka et al.(2004)]{tan04}
	Tanaka, M., Goto, T., Shimasaku, K., Okamura, S.,
	Shimasaku, K., \& Brinkmann, J.
	2004, \apj, 128, 2677

 \bibitem[Tran et al.(2003)]{tra03}
	Tran, K. H., Franx, M., Illingworth, G., Kelson, D. D., \& van Dokkum, P.
	2003, \apj, 599, 865

 \bibitem[Tran et al.(2004)]{tra04}
        Tran, K. H., Franx, M., Illingworth, G., van Dokkum, P., 
        Kelson, D. D., \& Magee, D.
        2004, \apj, 609, 683

 \bibitem[Yagi \& Goto(2006)]{yag06a}
	Yagi, M., \& Goto, T.
	2006, \aj, 131, 2050

 \bibitem[Yagi, Goto \& Hattori(2006)]{yag06b}
	Yagi, M., Goto, T., \& Hattori, T.
	2006, \apj, 642, 152


 \bibitem[Yamauchi \& Goto(2005)]{yam05}
	Yamauchi, C., \& Goto, T.
	2005, \mnras, 359, 1557


 \bibitem[Yang et al.(2004)]{yan04}
	Yang, Y., Zabludoff, D., Zaritsky, D.,
	Lauer, T., \& Mihos, J. C.
	2004, \apj, 607, 258

 \bibitem[Yasuda et al.(2001)]{yas01}
	Yasuda, N., et al. 2001, \aj, 122,1104

 \bibitem[York et al.(2000)]{yor00} 
	York, D. G., et al. 2000, \aj, 120, 1579
 

\end{thebibliography}
\end{document}